\documentclass[12pt,epsf]{article}

\usepackage[dvips]{graphicx}

\newcommand{\bnabla}{\mbox{\boldmath $\nabla$}}
\newcommand{\beq}{\begin{equation}}
\newcommand{\eeq}{\end{equation}}
\newcommand{\bea}{\begin{eqnarray}}
\newcommand{\eea}{\end{eqnarray}}

\begin{document}
\thispagestyle{empty}
\vspace*{-15mm}
\baselineskip 1pt
\begin{flushright}
\begin{tabular}{l}
{\bf YITP-04-28}\\
{\bf OCHA-PP-228}\\
{\bf May 2004}\\
{\bf hep-ph/0405259}
\end{tabular}
\end{flushright}
\baselineskip 24pt
\vglue 10mm

\begin{center}
{\LARGE\bf
Pentaquark Baryons in String Theory
}
\vspace{7mm}

\baselineskip 18pt
{\bf MASAKO BANDO$^1$, TAICHIRO KUGO$^2$, AKIO SUGAMOTO$^{3,4}$ and SACHIKO TERUNUMA$^4$}
\vspace{2mm}

{\it
   $^1$Physics Division, Aichi University, Aichi 470-0296, Japan\\ 
  $^2$Yukawa Institute, Kyoto University, Kyoto 606-8502, Japan\\
$^3$Department of Physics, Ochanomizu University, Tokyo 112-8610, Japan\\
$^4$Graduate School of Humanities and Sciences, Ochanomizu University, Tokyo 112-8610, Japan
}\\
\vspace{10mm}
\end{center}
\begin{center}
{\bf Abstract}\\[7mm]
\begin{minipage}{14cm}
\baselineskip 16pt
\noindent

The recently observed pentaquark baryons $\Theta^{+}$ and $\Xi^{--}$ are 
studied in the dual gravity theory of QCD.  By developing a general formulation useful for studying the branched string web in a curved space, simple mass formulae of the pentaquark baryons are derived in the Maldacena prototype model for  supersymmetric QCD and a more realistic model for  ordinary QCD.  Even though the approximations adopted here in deriving these formulae are extremely naive, the predicted values of the masses do not differ significantly from their experimental values.  We thus conclude that this approach is promising for the purpose of obtaining a description of the observed masses of pentaquarks and of their extremely narrow decay widths.  With the aim of constructing more reliable mass formulae, a preliminary sketch is given of how spin, the hyperfine interaction and parity are considered in the string picture.

\end{minipage}
\end{center}

\newpage
\baselineskip 18pt
\def\thefootnote{\fnsymbol{footnote}}
\setcounter{footnote}{0}

\section{Introduction}
   Even at the time that the quark model was introduced in the 1960s, Gell-Mann considered various exotic quark states in addition to ordinary mesons and triquark baryons~\cite{G}.  However, such exotic states were not observed for four decades.   For this reason, it made a great impact on us when the first exotic state, the pentaquark baryon $\Theta^{+}$, was discovered at SPring-8~\cite{SP8} last year.   To produce $\Theta^{+}$, a  $\gamma$ beam was generated by applying a 3.5 eV laser beam to an 8 GeV electron beam at SPring-8 through Compton backward scattering.  The induced process is $\gamma+n\rightarrow K^{-}\Theta^{+}\rightarrow K^{-}K^{+}n$, 
and then  $\Theta^{+}$
is an exotic pentaquark baryon consisting of five quarks, explicitly, $(ud)(ud)\bar{s}$.  The observed mass is rather light, $M(\Theta^{+})=1,540 \pm 
10 \mbox{MeV}$, and the width is extremely narrow, $\Gamma(\Theta^{+}) \le 25 \mbox{MeV}$.  Subsequently, 
another
pentaquark state, $\Xi^{--}((ds)(ds)\bar{u})$ was observed at CERN 
NA49~\cite{NA49} with mass $M(\Xi^{--})=1,862 \mbox{MeV}$, and width 
$\Gamma(\Xi^{--}) \le 18
\mbox{MeV}$.  

Prior to these experiments, these pentaquarks were predicted by
Diakonov {\it et al.}~\cite{DPP} as chiral solitons.  Their pioneering work motivated the experiment at SPring-8, which proved their prediction of the mass $M(\Theta^{+})=1,530 \mbox{MeV}$ and the width  $\Gamma(\Theta^{+}) 
\le 15 \mbox{MeV}$ to be consistent with the experimented values.  However, the chiral soliton prediction of the mass and width of the pentaquark $\Xi^{--}$ are outside the bounds of the NA49 experimented values.   Jaffe and Wilczek~\cite{JW} treated the penatquarks as being composed of two diquark pairs and one anti-quark, obtaining a good prediction for their masses.  In this picture, two quarks are correlated to form a diquark, and the two diquarks form a P-wave state, so that the (spin, parity) of $\Theta^{+}$ and $\Xi^{--}$ is $(\frac{1}{2}, +)$, instead of $(\frac{1}{2},-)$, and the representaion of the pentaquarks is 
$\overline{\bf 10}$ under flavor $SU(3)_{f}$
\footnote{
A good theoretical review of pentaquark baryons is given in~\cite{Oka}.
}.  This can be understood as follows. Such a diquark pair $(ud)$ in an S-wave state carries the same color and flavor 
quantum numbers as an anti-$s$-quark, and therefore we can consider it an anti-scalar $s$-quark and denote $\bar{\tilde{s}}$. Then, $\Theta^{+}$ can be represented as 
$\bar{\tilde{s}}_1\bar{\tilde{s}}_2 \bar{s}_3$, and it belongs to 
$\overline{\bf10}$ of $SU(3)_{f}$, corresponding to the anti-$\Omega^{-}$ of 
$\bar{s}_1\bar{s}_2 \bar{s}_3$.  
For this reason, we are tempted to consider the broken ``supersymmetry" between an anti-diquark bound state and an elementary quark state, namely $(\bar{u}\bar{d})\approx \tilde{s} \leftrightarrow s$.  We do not develop this idea further here.  Another interesting model was proposed by Karliner and Lipkin, in which the pentaquarks are made of two clusters, diquark $(ud)$ and triquark $(ud\bar s)$. These clusters form a P-wave, giving also $(\frac{1}{2}, +)$ and $\overline{\bf 10}$ under flavor $SU(3)_{f}$ for pentaquarks~\cite{KL}. The pentaquarks were also studied in lattice QCD~\cite{LAT}.

The purpose of this paper is to propose another picture of the pentaquark baryons in colored string theory. The best way to understand this picture is to draw a picture of 
$\Theta^{+}$ 
as quarks connected by colored strings of three colors, red, green and blue.  In this picture, a colored string (a colored flux) emerges  from a quark and terminates at the anti-quark with the opposite color to the initial quark, and three strings with different colors  (red, green and blue) are joined at a junction.  In this way, we have the very beautiful shape displayed in Fig.1.

\begin{figure}
\begin{center}
\includegraphics[width=10cm,clip]{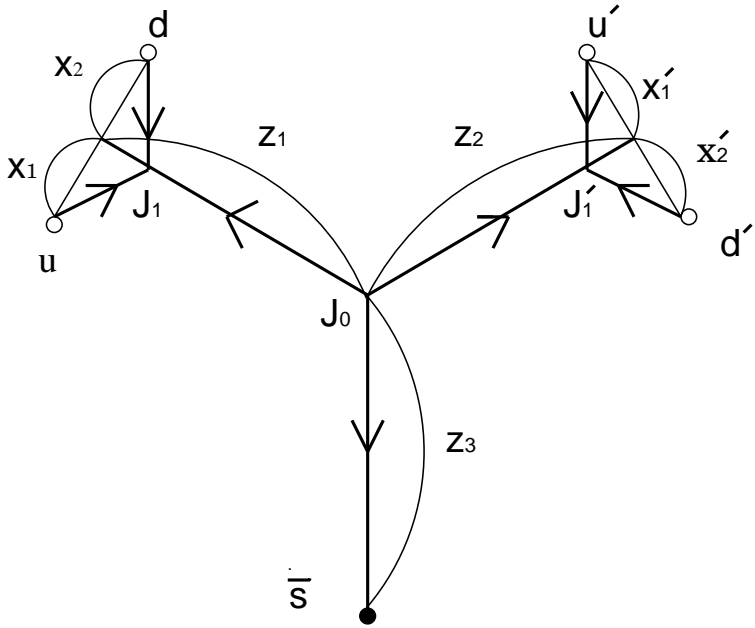}
\end{center}
\caption{Three-dimensional view of the pentaquark $\Theta^+$ in the string picture.}
\end{figure}%

In this picture, the mass of $\Theta^{+}$ is estimated as the total length of the colored strings.   Furthermore, there may be an unexpected merit of this picture; that is, this shape of the branched web is quite stable.  This is because, for a pentaquark to decay into a meson and a triquark baryon, string configurations with a loop must appear in the intermediate stage of decay, but this may be a rare occurrence.
To produce colored strings, however, non-perturbative QCD is necessary. 

As shown by Maldacena~\cite{M}  in 1998, we can replace non-perturbative QCD by a classical gravity theory, whitch is dual to QCD.  The space determined by this gravity theory is flat in the ordinary four dimensions but it is deformed in the extra dimensions.   
For this reason, we evaluate the mass of pentaquarks from the total length of the strings that are located in the curved background space with extra dimensions, as determined in the dual gravity theory of QCD.

In the manner described above, we obtain the mass formulae of pentaquark baryons.  In this paper we examine two different models.  One is the original prototype model of Maldacena~\cite{M} which possesses supersymmetry and is not QCD-like, but is the best context within which to examine our formulation.  The other is a model in which the supersymmetry is broken by compactifying one spatial dimension, following Witten's idea ~\cite{N=0a, N=0b, N=0d, N=0c}. \footnote{Preliminary results of this paper were reported in Ref.~\cite{S}.}  

In the next section, we review the dual gravity model of QCD in terms of the familiar comcepts of  factorization and the vacuum insertion.
In Sec.3 we present our general formulation for studying pentaquark baryons.  Application of this formulation to Mardacena's prototype model and a QCD-like model is given in Sec.4 and Sec.5, respectively.  In Sec.6 we study the formation and decay of pentaquark baryons. A rough sketch is given in Sec.7 of how spin, the hyperfine interaction and parity are understood in our string picture.  The last section is devoted to discussion.  In the appendix several mathematical formulae used in this paper are proved.

\section{Review-dual gravity model of QCD in terms of factorization and the vacuum insertion} 
As is well known, QCD consists of quarks and gluons.  Quarks possess both color $(r, g, b)$ and flavor $(u, d, s, {\it etc.})$, while gluons possess color $(r, g, b)$ and anti-color $(\bar{r}, \bar{g}, \bar{b})$ but not flavor. An open string (a string with two endpoints) is ideally suited to account for such quantum numbers at its two ends.  For quarks, one end represents color and the other end flavor.  For gluons, one end represents color and the other anti-color.  

In recently developed string theory, there are  "branes" (higher dimensional extended objects that are generalized membranes) to which the endpoints of an open string are confined . Applying this idea to QCD, we introduce $N_{c} (=3)$ ``colored branes" and $N_{f}$ ``flavored branes" at which open strings 
corresponding to quarks and gluons terminate.  In Fig. 2, we depict a gluon emission process from a u-quark: $u_{r} \rightarrow u_{b}+ g_{r\bar{b}}$.  
It is interesting that such a simple splitting of a string can represent the gauge interaction. 

\begin{figure}
\begin{center}
\includegraphics[width=10cm,clip]{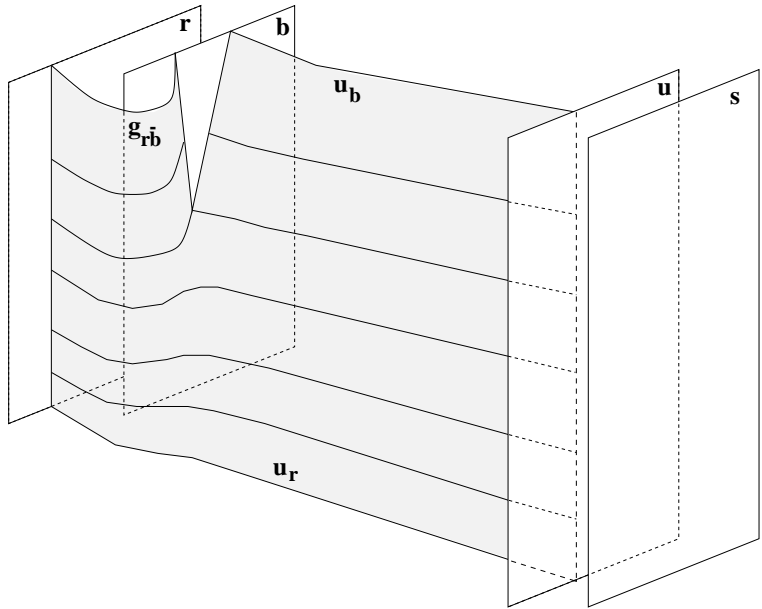}
\end{center}
\caption{Gluon emission of a quark, $u_r \to u_b + g_{r \bar{b}}$, in which the endpoints of the quarks and the gluon are confined and move on either the colored sheets $r$ and $b$ or the flavored sheet $u$.}
\end{figure}%

The energy of a string is given by the sum of the classical energy stored inside the  string and the excitation energies of vibration and rotation. 
Because the classical energy of a string is proportional to its length and because  
gluons are massless, $N_c$ colored branes should lie on top 
of one another. On the other hand, quarks possess intrinsic masses, 
and therefore the endpoints of a quark string, namely, a flavored brane and a colored
brane should be separated from each other by a nonvanishing distance $U$. Then, the intrinsic quark mass $m_q$ can be represented as $m_q=U\times 
\mbox{(string tension)} =U/2\pi\alpha'$, where the string tension is the 
energy stored inside a unit length of string and is represented in terms of $\alpha'$, historically called the Regge slope. 

To describe QCD, we have to prepare Dp-branes and Dq-branes with $p, q \ge 3$ for colored branes and flavored branes, respectively, and these branes should be located in the space of more than five dimensions.
\footnote{The reason is as follows. The motion of an open string is constrained ; that is, its endpoints cannot move in directions normal to the branes.  The boundary conditions on such endpoints are
called ``Dirichlet". By contrast, the endpoints can move freely in directions along the brane, which gives Neumann boundary conditions.  The name ``D-brane" is taken from these boundary condition. If the D-brane is a $p$-dimensionally extended object, we call it Dp-brane.  For QCD, we need three spatial dimensions on which quarks and gluons move freely, and therefore we have to prepare colored Dp-branes and flavored Dq-branes with $p, q \ge 3$.  To represent the intrinsic masses of quarks we need to introduce at least one extra dimension $u$ to realize the separation of the flavored branes from the colored branes, giving nonvanishing masses of quarks.  Now, the dimensionality of the space to accomodate branes should be greater than five. }

To evaluate the amplitude for a certain process to occur in the above picture, we have to sum up all the possible two-dimensional world sheets 
with the weight $\exp(iS)$, where the action $S$ is given by $S$=(energy)$\times$(time)=(area of the string's world sheet)/$2\pi\alpha'$, following the Feymann  path integral formulation.  As stated above, the endpoints of the strings are confined to the colored branes or flavored branes, so that the world sheet has boundary trajectories $\{C_{i}\} (i=1, 2, ...)$ on colored branes and $\{F_{j}\} (j=1, 2, ...)$ on flavored branes; that is, the  amplitude $A$ can be written 
\begin{equation}
A(\{C_{i}\}, \{F_{j}\})=\langle\{C_{i}\}, \{F_{j}\}\rangle, \label{original amplitude}
\end{equation}
where $\langle \rangle$ represents the weighted average taken with fixed boundary trajectories.   

With the situation as described above, let us make a simple approximation using factorization and a vacuum insertion, which is frequently used in ordinary high energy physics including QCD. 
\footnote{For example, in the semi-leptonic decay of $B^{0}\rightarrow K^{-}e^{+}\nu$, we factorize the current-current interaction and use the vacuum insertion $|vac\rangle\langle vac|$:  
$$\langle B^{0}|J^{+\mu}J^{-}_{\mu}|K^{-}, e^{+}, \nu\rangle\approx\langle B^{0}, K^{+} |J^{+\mu}| vac\rangle\langle vac|J^{-}_{\mu}|e^{+}, \nu\rangle.$$
The dual gravitational description of QCD can be understood in the same way as in ordinary QCD.}  
Explicitly, the amplitude in Eq. (\ref{original amplitude}) is approximated by the factorized amplitude with a vacuum insertion:
\begin{equation}
A(\{C_{i}\}, \{F_{j}\})\approx \langle\{C_{i}\}|vac\rangle\langle vac|\{F_{j}\}\rangle.
\end{equation}
The case of estimation of the potential between ${\bar u}$- and $s$-quarks is depicted in Fig. 3.
After summing up all the possible configuration of $\{C_{i}\} (i=1, 2, ...)$, we have 
\begin{eqnarray}
A(\{F_{j}\}) &\approx& \left( \sum_{\{C_{i}\}}\langle\{C_{i}\}|vac\rangle \right)\times \left(\langle vac|\{F_{j}\}\rangle\right) \\
&\propto&\langle vac|\{F_{j}\}\rangle.
\end{eqnarray}

\begin{figure}
\begin{center}
\includegraphics[width=10cm,clip]{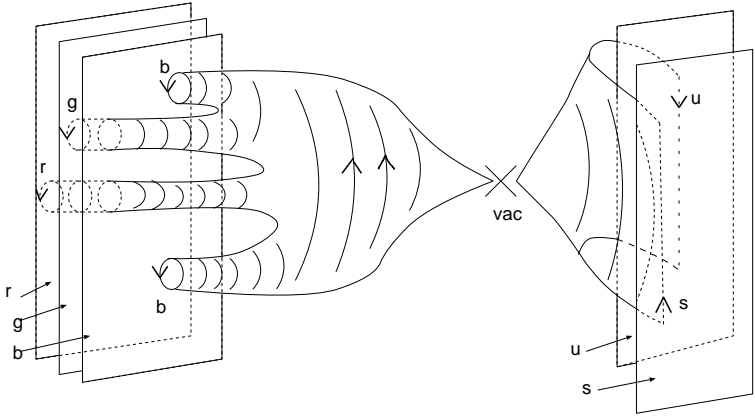}
\caption{ Factorization and vacuum insertion of the amplitude in string theory: an example of estimating the potential between $\bar{u}$ and $s$.}
\end{center}
\end{figure}%

Now, the remaining problem is to determine what the vacuum state is.  As seen from the first factor, $\sum_{\{C_{i}\}} \langle\{C_{i}\}|vac\rangle$, or from Fig.3, here, the existence of $N_{c}$ colored branes affects the vacuum state through the interactions mediated by closed strings.  In string theory, massless modes of closed strings are generally the graviton, an anti-symmetric tensor (Kalb-Ramond) field and a dilaton, which belong to the so-called Neveu Schwarz-Neveu Schwarz (NS-NS) sector, and if the theory is superstring, we have, furthermore, the fields belonging to the Ramond-Ramond (R-R) sector.  These (NS-NS) and (R-R) sectors describe bosonic fields, while the fermion fields come from the mixed (R-NS) and (NS-R) sectors.~\cite{String} The vacuum state is now represented as the classical configurations of the bosonic closed string modes of (NS-NS) and (R-R) when the $N_{c}$ colored branes are located in the space with extra dimensions.  
This is the method to determine the vacuum. To compensate for the error introduced by the approximation using the factorization and the vacuum insertion is usually very difficult.  
 \footnote{Sufficient conditions for the approximation to be valid are known to suppress the oscillation modes of strings by a factor of $\alpha' \rightarrow 0$ compared with the typical length squared in the theory, $R^2$, and to suppress the loop effects of strings by a factor of $g_{s}\rightarrow0$.  In the simplest case, they become the QCD coupling $\alpha_c
\to 0, N_c \to \infty$ and $N_c\alpha_c \to \infty$.  In the ordinary QCD, the improvement of the approximation based on factorization with a vacuum insertion has been carried out very recently~\cite{BBNS}.}

Next, we explain how the vacuum is determined within this approximation.  The important point is that the vacuum obtained here is not a usual flat space, but a curved space with compactification.  

A Dp-brane is known to have tension $\tau_{p}$ (the energy per unit volume), given by 
\begin{equation}
\tau_{p}^2=\frac{2\pi}{2\kappa^2}(4\pi^{2} \alpha')^{3-p}, \label{Dp-tension}
\end{equation}
where $\kappa$ is related to the ten-dimensional Newton constant $G_{10}$ or the closed string's three point coupling $g_{s}$ by~\cite{String}
\footnote{Equation(\ref{Dp-tension}) is derived by exploiting the fact that the gravitational potential between two Dp-branes can be calculated by two methods.  One method is the usual estimation of the Newton potential, giving a result proportional to $(\mbox{mass})^{2} \propto(\tau_{p})^{2}$.  The other is the method employing the closed string's exchange process between two Dp-branes, as was done by Polchinski~\cite{P}.  The results obtained using two methods are equated, and the result is Eq.(\ref{Dp-tension}).  Of course, in applying the second method we have to fix the string theory.  The above result is that obtained for a superstring of type IIA or type IIB, defined in ten space-time dimensions.~\cite{String}}
\begin{equation} 
2\kappa^2=16\pi G_{10}=(2\pi)^7 (\alpha')^4 (g_{s})^2.
\end{equation}  
Then, the existence of the  $N_{c}$ stuck Dp-branes with energy density $N_{c}\tau_{p}$ deforms the flat ten-dimensional space-time into a certain curved space-time with the non-flat metric $G_{\mu\nu}(x)$.  To understand the space-time so deformed, we have to follow the low energy supergravity theory effectively describing the superstring at low energy~\cite{HS}.  In these superstring theories, the Dp-brane has a charge $\mu_{p}$ for the antisymmetric tensor field with $(p+1)$ indices, or the Ramond-Ramond $(p+1)$-form fields. The charge is given by \begin{equation}
\mu_{p}^2=2\pi(4\pi^{2} \alpha')^{3-p}.
\end{equation}
Using the field strength with $(p+2)$ indices produced by this charge, the compactification into $S^{p+2}$ becomes possible~\cite{F}.  
If we prepare  $N_c$ D3-branes $(p=3)$ for the colored branes, $S^5$ compactification is realized.  The remaining five-dimensional space has been determined by Maldacena using the black 3-brane solution in the supergravity theory~\cite{HS}.  Taking a certain limit, he obtained the five-dimensional Anti de Sitter space, $AdS_{5}$.  Therefore, the background space in the presence of stacked $N_{c}$ D3 branes becomes $AdS_{5} \times S^{5}$. 
\footnote{$AdS_5$ is the space described by $-x_{-1}^2-x_0^2+x_1^2+x_2^2+x_3^2+x_4^2=R^2$ with a radius $R$ of the AdS space. Therefore, this space possesses the conformal symmetry $SO(2, 4)$, and $S^5$ gives the additional $SO(6)=SU(4)$ symmetry.  This symmetry, $SO(2, 4) \times SU(4)$, in the dual gravity theory corresponds exactly to that in the $N=4$ supersymmetric Yang-Mills theory, which is conformally invariant (i.e. the coupling does not run), possesses the $SU$(4) invariance, which is the symmetry of the $N=4$ generators of supersymmetries.}
  
In this manner, Maldacena's prototype model~\cite{M} of ``QCD" can be understood in terms of factorization and a vacuum insertion. In the model, supersymmetries remain maximally $N=4$, so that the corresponding gauge theory on the $N_{c}$ D3 branes becomes $N=4$ supersymmetric $SU(N_{c})$ Yang-Mills theory (gluodynamics) and is not ordinary QCD, even for $N_{c}=3$.

To describe the five-dimensional space, we have to introduce an extra coordinate $u$, in addition to the Minkowski space, $(t, z, {\bf x}_{\perp})$, along which the world volume of the stuck $N_{c}$ colored branes extend.  The $AdS_5\times S^5$ space is characterized by the following distance squared:
\begin{equation}
ds^2=f(u)(-dt^2+dz^2+d{\bf x}_{\perp}^2)+g(u)(du^2+u^2 d\Omega^2_5),
\end{equation}
where 
\begin{equation}
f(u)=g(u)^{-1}=(u/R)^2, \label{prototype model}
\end{equation}
and $d\Omega^2_5$ is the distance squared of $S^5$ with a unit radius.  The radius (typical length) of $AdS_5$ space $R$ is given by $g_{s}$, or the QCD coupling $\alpha_{c}=g_{c}^{2}/(4\pi)$, as follows:
\begin{equation}
R^4=4\pi g_{s}N_{c}=8\pi\alpha_{c} N_{c}. \label{R^4}
\end{equation} 

Here and hereafter, every variable is made dimensionless by multiplying proper powers of $\alpha'$ with the dimension
of $(\mbox{length})^2$.

In more realistic
models, we need to break the supersymmetries.  For this purpose, a very useful method is to compactify one dimension to a circle, following Witten~\cite{N=0a}.  If this
compactified dimension is time-like, we have a finite temperature theory with the Schwartzshild metric for a blackhole, and if the dimension is space-like, we have an extra $S^{1}$ in addition to the Minkowski space.  These can be called, in general, AdS Schwartzshild spaces~\cite{N=0a, N=0b}.  In this case, the boundary condition for the fermion field in the compactified dimension (time-like or space-like) should be anti-symmetric, while that of the boson field should be symmetric.  Therefore, if the radius of the Kaluza-Klein (K-K) compactified dimension is $R_{KK}$, then in terms of the mass scale $M_{KK}=2\pi/R_{KK}$, the mass spectrum of fermion and boson are, respectively, 
\begin{equation}
(m_{\rm fermion}, m_{\rm boson})
=M_{KK}\times(\mbox{half-integer, integer}),  \label{MKK}
\end{equation}
so that fermions (gluinos) become massive, with the gluons remaining massless, and the supersymmetries are broken completely.

In this paper, we choose the compactified dimension $\vartheta$ to be space-like, and we prepare a stack of $N_{c}$ D4-branes extending along the directions of the Minkowski space and the compactified direction $\vartheta$, following Ref.~\cite{N=0c}.  Then, the background metric of the dual QCD is described by
\begin{equation}
ds^2=f(u)(-dt^2+dz^2+d{\bf x}_{\perp}^2)+g(u)du^2+g(u)^{-1}d\vartheta^2+f(u)^{-1}u^2 d\Omega^2_4,
\end{equation}
where 
\begin{eqnarray}
f(u)&=&(u/R')^{3/2},~~g(u)=(f(u)h(u))^{-1}, \nonumber \\
h(u)&=&1-(U_{KK}/u)^3,  \label{QCD like model}
\end{eqnarray}
and $d\Omega^2_4$ is the distance squared of $S^4$ with unit radius.  The values of $R'$ and $U_{KK}$ are given by
\begin{equation}
R'^{3}=2\pi\alpha_{c}N_{c}/M_{KK},~~\mbox{and}~~ U_{KK}=\frac{8\pi}{9}\alpha_{c}N_{c}M_{KK}.
\end{equation}

As discussed above, we have to introduce $N_{f}$ flavored branes to introduce quarks with flavor.  Usually, in Maldacena's prototype model, the flavored branes are chosen as D7-branes extending along the Minkowski space plus four extra directions other than the $u$ direction~\cite{KK}.  The main reason to introducing the D7-brane as the flavored brane is that the open strings connecting the colored D3-branes and the flavored D7-brane give a hypermultiplet $(\mbox{quark}_{L,R}$, $\mbox{squark}_{L, R})$ of $N=2$ supersymmetric QCD belonging to the fundamental representation.  The other reason is the stability of the brane configuration.  It is well known~\cite{P} that no force acts between two parallel Dp-branes, because the energy density $\tau_{p}$ and the charge $\mu_{p}$ are equivalent, and the attractive force coupled to the energy through the exchange of gravitons in the (NS-NS) sector is canceled exactly by the repulsive force coupled to the charge through the exchange of the fields in the (R-R) sector.  Furthermore, it is known that no force acts between Dp and Dq branes when there are four mismatched directions along which the Dp-brane extends but the Dq-brane does not, or vice versa~\cite{String}.  Therefore, D3- and D7-branes can be stationary .  Similarly, in the more realistic model, the flavored branes are taken as D6-branes~\cite{N=0c}, extending in the directions including the Minkowski space and three extra directions, without including the compactified directions of $\vartheta $ and $u$.  As in the former case, the number of mismatched directions between the D4-brane and the D6-brane is four, and they can coexist in a stationary configuration.  
\footnote{In this respect, it may be possible to consider a non-stationary configuration of $N_{c}$ colored branes and $N_{f}$ flavored branes in the expanding universe.  With this choice, the freedom of choosing brane types may increase, and the masses of quarks, or the distances between colored branes and flavored branes, may change with the configuration  of the branes in time.  For example, it may be possible to account for the heaviness of $t$-quark by the rapid receding of the $t$-flavored brane from the stack of colored branes, even if all the colored and flavored branes start from the same big-bang point.}

\section{General formulation of Pentaquarks}

In this section we present a general formulation of pentaquark 
baryons in string theory.  As discussed in the previous section, to 
evaluate the mass of hadrons, such as pentaquarks, we have to evaluate the 
action of the world sheet of the hadron (or pentaquark) that is connected by 
colored strings located in the curved background space determined by the dual 
gravity theory of QCD, where the boundary trajectories of the string world sheet are fixed on the flavored branes. Specifically, we have to evaluate the amplitude
\begin{equation}
\langle vac|\{F_{j}\}\rangle.
\end{equation}
For this purpose, we start from the  string action in the background curved space, 
\begin{equation}
S= \frac{1}{2\pi} \int d\tau d\sigma 
\sqrt{-(\dot{X}^{M}\dot{X}_{M})(X'^{N}X'_{N})+(\dot{X}^{M}X'_{M})^2}, \label{action}
\end{equation}
where $X^{M}(\tau,\sigma)$ describes the configuration of the world sheet of a string with two parameters, $\tau$ and $\sigma$, and ~$\dot{}$~ and ~$'$~ are the derivatives  with respect to $\tau$ and $\sigma$, respectively.  Here, the outside space ({\it i.e.} the target space) of $X^{M}$ is curved, and the contraction $\dot{X}^{M}\dot{X}_{M}$, {\it etc.}, represents that with the metric $G_{MN}(x)\dot{X}^{M}\dot{X}^{N}$, {\it etc.}

To describe a background vacuum state (curved space) of the dual gravity theory of QCD, a simple choice of the distance squared, applicable to both the prototype model and the more realistic model, is the following: 
\begin{equation}
ds^2=f(u)(-dt^2+dz^2+d{\bf x}_{\perp}^2)+g(u)du^2. \label{metric}
\end{equation}
The five-dimensional metric $G_{MN}(x)$ can be read from this form.

By choosing $\sigma=z$ and $\tau=t$, we can fix the parameterization of the world sheet.  Then, in the static limit, $\dot{X}^{M}\rightarrow0$ for $M\ne0$, the string action (\ref{action}) in the time interval $\Delta t$ reads 
\begin{equation}
S=\frac{\Delta t}{2\pi}\int dz L, 
\end{equation}
with the ``Lagrangian"
\begin{equation}
L=\sqrt{f(u)^2(1+({\bf x}'_{\perp})^2))+f(u)g(u)(u')^2}. \label{L}
\end{equation}

If we regard $z$ as ``time", there appear three conserved quantities, ``energy" 
$H$, and the two-component perpendicular ``momentum", ${\bf p}_{\perp}$.  The ``energy" is conserved, because the ``Lagrangian" Eq.(\ref{L}) does not depend on $z$ explicitly.  In order to conserve the energy we use Cartesian coordinates instead of polar coordinates.  The momentum ${\bf p}_{\perp}$ is conserved because its ``conjugate coordinates" ${\bf x}_{\perp}$ do not appear in the ``Lagrangian".  More explicitly, we have
\begin{equation}
(-H)=\frac{f(u)^2}{L}, ~~\mbox{and}~~ ({\bf p}_{\perp})={\bf x}'_{\perp} (-H). \label{pperp}
\end{equation}
Then, the perpendicular 
coordinates move uniformly in ``time", i.e.,
\begin{equation}
{\bf x}_{\perp}=\frac{({\bf p}_{\perp})}{(-H)}z. \label{xperp}
\end{equation}  
From Eqs.(\ref{pperp}) and (\ref{xperp}), we have
\begin{equation}
u'=\frac{1}{(-H)} \sqrt{\frac{f}{g}\left( f^2-(-H)^2-({\bf p}_{\perp})^2 \right)}. 
\label{eq:shape}
\end{equation}

Now, the ordinary energy stored inside a string, or the ``action" per unit time, during the change of $u$ from ${U_1}$ to ${U_2}$, can be estimated as
\begin{eqnarray}
E&=&\frac{1}{2\pi}\int dz L =\frac{1}{2\pi}\int du \frac{1}{u'} \frac{f(u)^2}{(-H)} \\
&=& \frac{1}{2\pi} \int^{U_2}_{U_1}du \sqrt{ 
\frac{f(u)^3g(u)}{f(u)^2-(-H)^2-({\bf p}_{\perp})^2}}. \label{Energy}
\end{eqnarray}

We can also estimate the change of $z$ from ${U_1}$ to ${U_2}$ as
\begin{equation}
z=(-H) \int^{U_2}_{U_1}du~~\sqrt{ 
\frac{g(u)}{f(u) \left(f(u)^2-(-H)^2-({\bf p}_{\perp})^2 \right)}}, \label{z}
\end{equation}
from which the change of ${\bf x}_{\perp}$ can be estimated with the help of Eq.(\ref{xperp}).
The shape of the string is determined by Eq.~(\ref{eq:shape}).
 
As a warm up, let us first consider the shape of the string for the 
simplest case of a meson, $q\bar q$, where $q$ and $\bar q$ are 
placed on the flavored brane at $u=U_1$ separated by a distance 
$2r$. The string connecting this quark and anti-quark is clearly on 
a plane, which we take as the $(z, u)$-plane with ${\bf x}_{\perp}={\bf0}$, and it reaches a certain minimum value $U_0$ of $u$ at the mid-point which  
we take to be at $z=0$. 
Then, from the symmetry of the system, the string profile can be parametrized 
as $(\pm z, u=u(z))$, where the signs $\pm$ correspond to the half parts from the 
mid-point ($z=0$) between the quark ($z=r$) and anti-quark ($z=-r$), 
respectively. Because $u'\ge 0$, $u(z)$ increases monotonically 
from $u(0)=U_0$ to $u(r)=U_1$. 

The quantity $U_1$ was a fixed value, the position of the flavored plane. The minimum 
value $U_0$ is, on the other hand, determined by the separation 
parameter $r$ via Eq.~(\ref{z}). Now, we have ${\bf x}'_{\perp}={\bf0}$ 
and $(-H)=f(U_0)$, because $u'=0$ at the mid-point, and therefore 
Eq.~(\ref{z}) reads 
\begin{equation}
r= f(U_0)\int^{U_1}_{U_0} du~~\sqrt{ 
\frac{g(u)}{f(u) \left(f(u)^2-f(U_0)^2 \right)}}.
\end{equation}
With a fixed $U_1$, this can diverge only when $U_0$ approaches the 
singularity of $g(u)/f(u)$, which is at $u=0$ and $u=U_{KK}$ for Maldacena's prototype metric case (\ref{prototype model}) and 
the QCD-like model case (\ref{QCD like model}), respectively. [Note that, 
for the latter case, $u$ must satisfy the relation $u\ge U_{KK} (>0)$, and therefore $u=0$ cannot be realized.] We refer to these singular values  $g(u)/f(u)$ generically as $U_{\rm MIN}$. Therefore, for the 
large separation limit $r \to \infty$, the minimum $u=U_0$ approaches 
the singularity $U_{\rm MIN}$. 

The energy expression (\ref{Energy}) can be written, by using (\ref{L}), as
\begin{equation}
E=\frac{1}{2\pi}\int dz L 
= \frac{1}{2\pi} \int^{U_1}_{U_0} \sqrt{ 
f(u) \left(f(u)dz^2+g(u)du^2\right)}. 
\end{equation}
This consists of two contributions, the horizontal part, $f(u)dz$, and 
vertical part, $\sqrt{f(u)g(u)}du$. 
The string shape is determined in such a manner to minimize this 
energy for a given separation $2r$ between the quark and antiquark. 
Because $f(u)\propto u^p$ with a positive power $p$, $f(u)$ becomes 
smallest at $u=U_{\rm MIN}$. Therefore, a displacement along 
the horizontal direction (the $z$ direction) as near $u=U_{\rm MIN}$ as possible is energetically favored. 
In the large separation limit $r\to \infty$, therefore, 
the string departing from the quark at $u=U_1$ ($z=r$) quickly goes down 
vertically to the lowest possible value, $u=U_0\approx U_{\rm MIN}$, 
and then 
moves horizontally from $z=r$ to $z=-r$ nearly maintaining the relation  
$u=U_0\approx U_{\rm MIN}$, and then it goes up to the anti-quark at $u=U_1$.
Schematically, the shape of the string is depicted in Fig. 4.

\begin{figure}
\begin{center}
\includegraphics[width=10cm,clip]{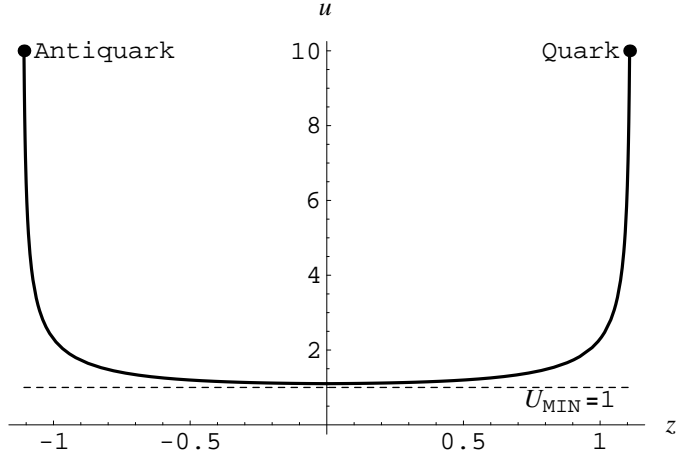}
\end{center}
\caption{Shape of the string connecting the quark and anti-quark in the QCD-like model.}
\end{figure}%

The energy of the vertical part and horizontal part of 
the half string are evaluated respectively as  
\begin{equation}
E_{\rm vertical} 
= \frac{1}{2\pi} \int^{U_1}_{U_0} \sqrt{f(u)g(u)}du 
\approx \frac{1}{2\pi} (U_1-U_{\rm MIN}) \equiv m_q,
\end{equation}
\begin{equation}
E_{\rm horizontal} 
= \frac{1}{2\pi} \int^{r}_{0} f(u) dz 
\approx \frac{1}{2\pi} f(U_{\rm MIN}) r.
\label{horizontal}
\end{equation}
We have defined the quark mass $m_q$ of flavor $q$ to be 
the energy of the string stretching along a line from the quark on the 
flavored brane to the minimum of $u$, $U_{\rm MIN}$. 
It is interesting that the linear potential $\propto r$ appears as long 
as $f(U_{\rm MIN})\not=0$, which the QCD-like case 
(\ref{QCD like model}). However, $f(U_{\rm MIN})=f(0)=0$ in 
Maldacena's prototype model (\ref{prototype model}), in which case a more 
careful calculation is necessary to evaluate the next leading term.

We now come back to the pentaquark system.
In Fig. 1 we can see three junctions  of strings, $J_0, J_1$ and $J_1'$.  At $J_0$, three strings are assumed to separate with opening angles $2\pi/3$ in the $u=U_{0}$ plane, satisfying $u'=0$ at $J_0$.  This splitting can be naturally understood.  At $J_1$ (or $J_1'$), string (1) coming from $J_0$, with coordinates 
$(z, {\bf x}_{\perp}={\bf 0}, u_{(1)}(z))$, is assumed to split into 
string (2) with coordinates $(z, {\bf x}_{\perp}, u_{(2)}(z))$ and string (3) with $(z, -{\bf x}_{\perp}, u_{(3)}(z))$; that is, string (1) stretching in the $z$ direction splits into strings (2) and (3), which depart from each other in the ${\bf x}_{\perp}$ direction.  

A five-dimensional view of $\Theta^{+}$ is shown in Fig. 5.

\begin{figure}
\begin{center}
\includegraphics[width=15cm,clip]{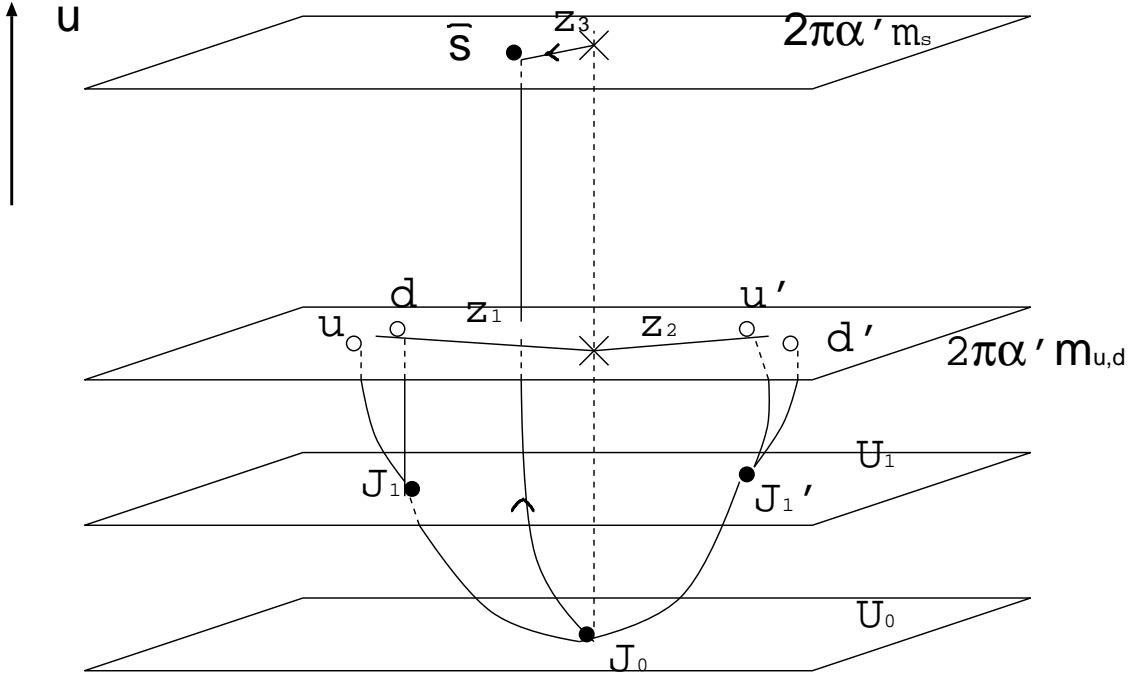}
\end{center}
\caption{Five-dimensional view of the pentaquark $\Theta^+$. Here the vertical axis corresponds to the extra dimension $u$.}
\end{figure}

In this situation, let us find the connection condition of the three strings at the junctions.
Employing a locally flat coordinate system near the junctions $J_1$ and $J_1'$, three 
strings lie in a single plane, pulling each other with equal tension.  As a result, the opening angles between strings are all $2\pi/3$, at least in the locally flat coordinate system on the plane.  

In this locally flat coordinate system, the vector in the direction tangential to the string reads
\begin{equation}
d\bar{X}^{M}=(\sqrt{f(u)}dX^{\mu}, \sqrt{g(u)}du)\propto(\sqrt{f(u)}X'^{\mu}, \sqrt{g(u)}u').
\end{equation} 
Now, the tangential vectors for strings (1), (2) and (3) are given, respectively, by
\begin{eqnarray}
{\bf v}_{(1)}&=&(\sqrt{f(u)}, {\bf 0}_{\perp}, \sqrt{g(u)}u'_{(1)}), \\
{\bf v}_{(2)}&=&(\sqrt{f(u)}, \sqrt{f(u)}{\bf x}'_{\perp}, \sqrt{g(u)}u'_{(2)}), \\
{\bf v}_{(3)}&=&(\sqrt{f(u)}, -\sqrt{f(u)}{\bf x}'_{\perp}, \sqrt{g(u)}u'_{(3)}).
\end{eqnarray}
Therefore, we can write down the condition that the angle is equal to $\pi/3$ between strings (2) [or (3)] and (1) and that the angle is $2\pi/3$ between strings (2) and (3) as follows:
\begin{equation}
\frac{({\bf v}_{(2),(3)} \cdot {\bf v}_{(1)})}{({\bf v}_{(2),(3)})^2 ({\bf v}_{(1)})^2}=\frac{1}{2}, ~~\frac{({\bf v}_{(2)} \cdot {\bf v}_{(3)})}{({\bf v}_{(2)})^2 ({\bf v}_{(3)})^2}=-\frac{1}{2}. 
\end{equation}
More explicitly, we have
\begin{eqnarray}
\frac{f(u)+g(u)(u'_{(2), (3)})(u'_{(1)})}{ \sqrt{f(u)+g(u)(u'_{(1)})^2} \sqrt{f(u)\left(1+({\bf x}_{\perp})^2_{(2), (3)}\right)+g(u)(u')^2_{(2), (3)}}}&=&\frac{1}{2}, \\
\frac{ f(u)\left(1-({\bf x}_{\perp})^2_{(2), (3)}\right)+g(u)(u')^2_{(2), (3)} }{ f(u)\left(1+({\bf x}_{\perp})^2_{(2), (3)}\right)+g(u)(u')^2_{(2), (3)} }&=&-\frac{1}{2}.
\end{eqnarray}
From these, we 
obtain the following connection conditions at the junction $J_1$ or $J_1'$:
\begin{eqnarray}
u'_{(2), (3)}&=&u'_{(1)}, \label{condition1} \\
({\bf x}'_{\perp})^2_{(2), (3)}&=&3\left(1+\frac{g(u)}{f(u)}(u'_{(1)})^2\right) \label{condition2}.
\end{eqnarray}
From these two conditions, we can derive the condition
\begin{equation}
(-H)_{(2)}=(-H)_{(3)}=\frac{1}{2} (-H)_{(1)}.  \label{condition3}
\end{equation}
Equation (\ref{condition1})-(\ref{condition3}) constitute the connection conditions at the junctions $J_1$ and $J_1'$.

The results obtained here imply that the conservation of ``energy" and ``perpendicular momenta" hold during the process in which the splitting of the strings occurs.  As for the condition on ${\bf x}_{\perp}$, we know that ${\bf x}'_{\perp}=\sqrt{3}$ implies that strings (2) and (3) split with an opening angle of $2\pi/3$ in Minkowski space, after projecting out the fifth coordinate $u$, and therefore the angle becomes generically larger than  $2\pi/3$ for a tilted configuration in five-dimensional space.

With the connection conditions derived in this section, it is straightforward to evaluate the static energy 
of pentaquarks in terms of the coordinates of the quarks.  To this end, we must eliminate from the expression of the energy $E$ in Eq. (\ref{Energy}), the undetermined values of $u,  U_{0}$ and $U_{1}$ at the junctions $J_0$ and $J_1$ (or $J_1'$), respectively, and express the energy in terms of the coordinates $z$ and ${\bf x}_{\perp}$ of the branches of the strings in the pentaquark as well as the quark masses, as was done in the first example considered above.  

This general formulation is also applicable to the more complicated web-like exotic hadrons and string systems.

\section{Penatquarks in Maldacena's prototype model}

First we apply our general formulation to the evaluation of the masses of pentaquarks in Maldacena's prototype model.
The pentaquarks consist of two parts. The first part is the sum of the segments of strings, $(J_0, J_1)$,  $(J_0, J_1')$ and $(J_0, \bar{s})$, if they are specified by a pair of endpoints.  Then, the calculation of the energy for this part is the same as that done previously by many people~\cite{Wilson op}. Here, it is helpful to recall the example of the $(q{\bar q})$ meson system considered in the last section.

As discussed in Sec.3, the conserved energy becomes 
\begin{equation}
(-H)^2_{(1)}=f(U_{0})^2, \label{(-H)1}
\end{equation}
which is understood from the condition of $u'=0$ at the junction $J_0$, where $u=U_0$. 
Using the functions $f(u)$ and $g(u)$ given in Eq. (\ref{prototype model}) and the conserved energy in Eq. (\ref{(-H)1}), the energy and the coordinate of each segment of the first part are given by
\begin{eqnarray}
E_1&=&\frac{U_0}{2\pi} \int^{U_1/U_0}_{1} dy \frac{y^2}{\sqrt{y^4-1}},  \label{E_1} \\
z_{(1)}&=&\frac{R^2}{U_0} \int^{U_1/U_0}_{1} dy \frac{1}{y^2\sqrt{y^4-1}}, \label{z_1}
\end{eqnarray}
where  $y=u/U_0$, and $U_1$ is the value of $u$ at the other endpoint of the segments. 

Here we adopt the simplest 
approximation in which $U$ having a large extent, i.e., $U_0\ll 
U_1$.  Then, Eq. (\ref{z_1}) gives \begin{eqnarray}
U_0&=&\frac{c_1 R^2}{z_{(1)}}+\cdots, \\
c_1&=&\int^{\infty}_{1} dy \frac{1}{y^2\sqrt{y^4-1}}\approx0.599,
\end{eqnarray}
and thus, from Eq.(\ref{E_1}), we have
\begin{eqnarray}
E_1(U_1; z_{(1)})
&=&\frac{U_0}{2\pi} \left(
\int^{U_1/U_0}_{1}\!dy\,1 +
\int^{U_1/U_0}_{1} dy \biggl({y^2 \over\sqrt{y^4-1}}-1 \biggr)
\right) \nonumber\\
&=&
\frac{1}{2\pi} \left(U_1+(\tilde c_1-1)U_0+\cdots\right)
=\frac{1}{2\pi} \left(U_1-\frac{c_1^2 R^2}{z_{(1)}}+\cdots\right), \label{E_1'}
\end{eqnarray}
where
\begin{equation}
\tilde c_1 \equiv\int^{\infty}_{1} dy \left(
{y^2 \over\sqrt{y^4-1}}-1 \right), 
\end{equation}
with
\begin{equation}
c_1+\tilde c_1 =1. \label{A1} 
\end{equation}
This last relation is proved in the appendix.

Next, we analyze the second part, which consists of the diquark pairs $(u, d)$ and $(u', d')$.  The corresponding four segments of the string are  $(u, J_1), (d, J_1), (u', J_1')$ and $(d', J_1')$.
To evaluate the energy and the location of the segments, we choose the coordinate system discussed in the previous section, $(z, {\bf x}_{\perp}, u)$, for each segment. Then, from the connection condition Eq.(\ref{condition3}), we have
\begin{equation}
(-H)_{(2), (3)}=\frac{1}{2}f(U_0). \label{(-H)'}
\end{equation}
Replacing $(u'_{(1)})^2$ in Eq. (\ref{condition2}) by 
\begin{equation}
(u'_{(1)})^2=\frac{f(U_1)}{g(U_1)}\left(\left(\frac{f(U_1)}{f(U_0)}\right)^2-1\right),
\end{equation}
we have
\begin{equation}
({\bf x}'_{\perp})^{2}_{(2), (3)}=3\frac{f(U_1)^2}{f(U_0)^2}, \label{xperp2}
\end{equation}
which gives the following, by Eq. (\ref{pperp}): 
\begin{equation}
({\bf p}_{\perp})^2_{(2), (3)}=\frac{3}{4}f(U_1)^2. \label{(pperp)'}
\end{equation}

Then, we can evaluate the energy $E_2$ and the coordinate $z_2$ by using the general formulae Eq.(\ref{Energy}) and  Eq.(\ref{z}) as follows:
\begin{eqnarray}
E_2&=&\frac{U_1}{2\pi} \int^{U_2/U_1}_{1} dy \frac{y^2}{\sqrt{y^4-A^2}},  \label{E2} \\
z_{(2), (3)}&=&\frac{R^2U_0^2}{2U_1^3} \int^{U_2/U_1}_{1} dy \frac{1}{y^2\sqrt{y^4-A^2}}, \label{z2}
\end{eqnarray}
where $y=u/U_1$, and
\begin{equation}
A^2=\frac{\left( (-H)^2+({\bf p}_{\perp})^2 \right)_{(2), (3)}}{f(U_1)^2}=\frac{1}{4}\left(\frac{U_0}{U_1} \right)^4+\frac{3}{4}.
\end{equation}

Again, in the limit of large extent for $U$, i.e., $U_0 \ll U_1 \ll U_2$, we have $A^2 \approx 3/4$.  Accordingly, Eq. (\ref{z2}) gives
\begin{eqnarray}
U_1^3&=&\frac{c_2 R^2 U_0^2}{2z_{(2),(3)}}+\cdots, \label{U^3_1} \\ 
c_2&=&\int^{\infty}_{1} dy \frac{1}{y^2\sqrt{y^4-3/4}}\approx0.426.
\end{eqnarray}
In this case, it is better to use the transverse coordinate, ${\bf x}_{\perp(2), (3)}$, instead of the  longitudinal coordinate $z_{(2),(3)}$.  From Eqs. (\ref{xperp}), (\ref{(-H)'}) and (\ref{(pperp)'}), we obtain
\begin{eqnarray}
|{\bf x}_{\perp(2), (3)}|&=&\sqrt{3} \frac{f(U_1)}{f(U_0)} z_{(2), (3)}, \\
&=&\sqrt{3}\left(\frac{U_1}{U_0}\right)^2 z_{(2),(3)}.
\end{eqnarray}
This equation together with Eq. (\ref{U^3_1}) yields
\begin{equation}
U_1=\frac{\sqrt{3}}{2}c_2R^2\frac{1}{|{\bf x}_{\perp(2), (3)}|} .
\end{equation}

Now, from Eq. (\ref{E2}), we have 
\begin{eqnarray}
E_2(U_2; |{\bf x}_{\perp(2), (3)}|)&=&\frac{1}{2\pi}(U_2-c_3U_1) +\cdots \nonumber \\
&=&\frac{1}{2\pi} \left( U_2- \frac{\sqrt{3}}{2}c_2c_3R^2 \frac{1}{2} \left( \frac{1}{|{\bf x}_{\perp(2)}|}+\frac{1}{|{\bf x}_{\perp(3)}|} \right) \right) +\cdots, \label{E2'}
\end{eqnarray}
where 
\begin{equation}
c_3=1- \int^{\infty}_{1} dy \left( \frac{y^2}{\sqrt{y^4-3/4}}-1 \right)\approx 0.819,
\end{equation}
and the contribution from strings (2) and (3) is averaged.
The constants $c_2$ and $c_3$ are not independent, being related as
\begin{equation}
2c_3-1=\frac{3}{2} c_2.  \label{A2}
\end{equation}

Next, applying Eqs.(\ref{E_1'}) and (\ref{E2'}) to the first part 
and the second part of the pentaquark's segments, respectively, we obtain
\begin{eqnarray}
& & E(\mbox{pentaquark}:(q_1q_2)(q'_1q'_2)\bar{q}_3) \nonumber \\
&=&E_1(J_0, J_1)+E_1(J_0, J_1')+E_1(J_0, \bar{q}_3) \nonumber \\
&+&E_2(J_1, q_1)+E_2(J_1, q_2)+E_2(J_1', q'_1)+E_2(J_1', q'_2)  \\
&=&m_3+2(m_1+m_2)+V(z_{1,2,3}; x_{1,2}, x'_{1,2}), \label{defPotential}
\end{eqnarray}
where use has been made of the definition of the quark mass $m=U/2\pi$ with 
$u$-coordinate $U$ of the flavor brane on which the quark exists, and
\begin{equation}
V=-\frac{aR^2}{2\pi}\left(\frac{1}{z_1}+\frac{1}{z_2}+\frac{1}{z_3}\right)-\frac{bR^2}{2\pi} \frac{1}{2} \left(\frac{1}{x_1}+\frac{1}{x_2}+\frac{1}{x'_1}+\frac{1}{x'_2}\right),
\end{equation}
where $R$ is given in Eq.(\ref{R^4}).
Here, $z_{1}, z_{2}$  and  $z_{3}$ are, respectively, the distances
of $J_1$, $J_1'$ and $\bar{s}_3$ from $J_0$.  However, in the limit of large extent for $U$, $z_{(2), (3)} \ll|{\bf x}_{\perp(2), (3)}| \ll z_{(1)}$, the above defined $z_1$ and $z_2$ can be considered the center-of-mass coordinates of the diquark systems $(q_1q_2)$ and $(q'_1q'_2)$ from the junction $J_0$, while $|{\bf x}_{\perp(2), (3)}|$ becomes the internal coordinates $(x_{1}, x_{2})$ and $(x'_{1}, x'_{2})$ inside the diquarks $(q_1, q_2)$ and $(q'_1, q'_2)$, respectively. (See Figs.1 and 5 discribing $\Theta^{+}$)  For $\Theta^{+}=(ud)(u'd')\bar{s}$, $m_1=m_2$, so that $x_1=x_2$ and $x'_1=x'_2$ hold, and these are identically the relative coordinates of the diquark system.  For $\Xi^{--}=(ds)(d's')\bar{u}$, $m_1 \ne m_2$, and we thus find that $x_1 \ne x_2$ and $x'_1 \ne x'_2$.

The calculable constants $a$ and $b$ are given as 
\begin{equation}
a=c_1^2\approx0.359, ~~\mbox{and}~~b=\frac{\sqrt{3}}{2}c_2(2c_3-1)=\frac{\sqrt{3}}{4}c_2^2 \approx 0.236, 
\end{equation}
where Eq.(\ref{A2}) has been used, and the values of $a$ and $b$ are model dependent.

Next, we have to evaluate the mass eigenvalues of the pentaquark baryons.  In doing this we assume that the total Hamiltonian $\hat{H}$ is originally
\begin{equation}
\hat{H}=\sum_{i=1-5}\sqrt{(\hat{p}_{i})^2+m_{i}^2}+V(z_{1,2,3}; x_{1,2}, x'_{1,2}),
\end{equation}
from which the energy of the pentaquark is obtained as in the above discussion in the static limit.  To make this part more rigorous, the collective coordinates of the static pentaquark state should be analyzed carefully, following the general method employed in the soliton theory.  Here, without carrying out a careful analysis, we simply proceed as above.   Furthermore, we do not take into account motion that changes the angles between segments of strings, nor vibration modes of the string segments.  We take account of only the modes that change the lengths of string segments.  The modes changing the lengths of the ``center of mass" coordinates, $z_{1, 2, 3}$,  and those of the ``relative coordinates", ${\bf x}_{1, 2}$ and ${\bf x'}_{1, 2}$, are all independent.  The former modes give one-dimensional motion and the latter give two-dimensional transverse motion.

In this restricted treatment, the non-relativistic Sch\"odinger equations of our problem read 
\begin{eqnarray}
\left( \frac{-1}{2(m_3)}\left(\frac{\partial}{\partial z_3} \right)^2 + V_{z}(z_3) \right) \Psi(z_3)&=&\epsilon_3\Psi(z_3), \\
\left( \frac{-1}{2(m_1+m_2)}\left(\frac{\partial}{\partial z_{1,2}} \right)^2 + V_{z}(z_{1,2}) \right) \Psi(z_{1,2})&=&\epsilon_{1,2}\Psi(z_{1,2}), \\
\left(\frac{-1}{2m_i}\left( \left(\frac{\partial}{\partial x_i}\right)^2 +\frac{1}{x_i} \frac{\partial}{\partial x_i} \right)+ V_{\perp}(x_i) \right) \psi(x_i)&=&\varepsilon_{i\perp}\psi(x_i),
\end{eqnarray}
where the potentials are all of the Coulomb type,
\begin{equation}
V_{z}(z_i)=-\frac{aR^2}{2\pi} \frac{1}{z_i}, ~~\mbox{and}~~
V_{\perp}(x_i)=-\frac{1}{2}\frac{bR^2}{2\pi}\frac{1}{x_i}. 
\end{equation}

Because the lowest eigenvalue $E_0$ for the system of Coulomb potential 
$-A/r$ is known to be
\begin{equation}
E_0= -{1\over2}m \cases{
\displaystyle \left({2A\over2}\right)^2 & for $d=1$ dimension, \cr
\displaystyle \left({2A\over d-1}\right)^2 & for $d\ge2$ dimension, \cr}
\end{equation}
we have 
\begin{equation}
\epsilon_3=-\frac{1}{2}m_3\left(\frac{aR^2}{2\pi}\right)^2,~~~ \epsilon_{1,2}=-\frac{1}{2}(m_1+m_2)\left(\frac{aR^2}{2\pi}\right)^2  
\end{equation}
and
\begin{equation}
\varepsilon_{i\perp}=-\frac{1}{2}m_i\left(\frac{bR^2}{2\pi}\right)^2,
\end{equation}
where $R^4=8\pi\alpha_{c}N_{c}$, given in Eq.(\ref{R^4}).

Now, we obtain the mass of 
$\Theta^{+}$ as
\begin{equation}
M(\Theta^{+})=2(m_u+m_d)(A+B)+m_s A,
\end{equation}
where $A=1-\alpha_c N_c a^2/\pi$ and $B=-\alpha_c N_c b^2/\pi$.
Similarly, we obtain the mass of $\Xi^{--}$ as
\begin{equation}
M(\Xi^{--})=2(m_d+m_s)(A+B)+m_u A.
\end{equation}
In general, the mass formula of the pentaquark family of $\Theta^{+}$ is given by
 \begin{equation}
M((q_1q_2)(q'_1q'_2)\bar{q}_3)=2(m_1+m_2)(A+B)+m_3 A,
\end{equation}
while that of the triquark family of nucleons reads
\begin{equation}
M(q_1q_2q_3)=(m_1+m_2+m_3)A.
\end{equation}

Even though the prototype model and our approximation are very naive, if we employ the constituent quark masses $m_u=m_d=360  \mbox{MeV}$ and $m_s=540  \mbox{MeV}$ for the intrinsic quark masses and take $N_{c}=3$, then for the values of $\alpha_c$ needed to obtain the observed masses of pentaquarks and triquarks for $\Theta^{+}$ and $\Xi^{--}$ we have 1.38 and 0.79 for $\alpha_c$, while for the the triquarks, $N$, $\Sigma$ and $\Xi$, we have 1.07, 0.43, and 0.69, respectively.  These values exibit relations that are qualitatively similar to the observed value $\alpha_c=0.35 \pm$0.03 at $M_{\tau}=1,777 \mbox{MeV}$. We therefore conclude that  this approach to the study of pentaquarks is worth pursuing.

\section{Pentaquarks in a QCD like model}
In this section,  we consider a more realistic QCD-like model~\cite{N=0a, N=0b, N=0d, N=0c}.   In this model, we prepare $N_{c}$ colored D4 branes and $N_{f}$ flavored D6 branes, as discussed in Sec. 2.
The strings connecting the quarks of the pentaquark can stretch into extra dimensions outside the usual five dimensions, $(t, z, {\bf x}_{\perp}, u)$.  However, we do not consider this possibility here.  Then, we can apply the general formulation given in Sec.3 also to this QCD-like model.  Here, we must use $f(u)$ and $g(u)$ given in Eq. (\ref{QCD like model}).

For the three string segments,  $(J_0, J_1), (J_0, J_1')$, and $(J_0, \bar{q}_3)$ belonging to the first part, we have the following equations from Eqs. (\ref{Energy}) and (\ref{z}): 
\begin{eqnarray}
E_1&=&\frac{1}{2\pi} \int^{U_1}_{U_0} du~~  \frac{u^3}{\sqrt{ (u^3-U^3_{KK})(u^3-U^3_0)}},  \\
z_{(1)}&=&(U_0 R')^{3/2} \int^{U_1}_{U_0} du~~ \frac{1}{\sqrt{ (u^3-U^3_{KK})(u^3-U^3_0)} }.
\end{eqnarray}
If, again,  we consider the limit of large extent for $U$, i.e., $U_0 \ll U_1$, the minimum value of $u$, $U_0$, comes near the singularity 
at $u=U_{KK}$. Then we can rewrite the expression for $E_1$ as
\begin{equation}
E_1=\frac{1}{2\pi} \int^{U_1}_{U_0} du \,\sqrt{
\frac{u^3-U^3_{KK}}{u^3-U^3_0} } \ \ 
{}+ \frac{1}{2\pi}\left(\frac{U_{KK}^2}{U_0R'}\right)^{3/2} z_{(1)}.
\end{equation}
Despite the singularity at $u=U_0$, the integral of the first term 
can be approximated accurately by replacing the integrand by 1.
Thus we obtain the linear potentials for the three segments in the first 
part as
\begin{equation}
E_1=\frac{U_1}{2\pi} + k z_{(1)},
\end{equation}
with the coefficient
\begin{equation} 
k=\frac{1}{2\pi}\left(\frac{U_{KK}}{R'}\right)^{3/2}=(\frac{2}{3})^{3} \alpha_{c} N_{c} (M_{KK})^2.
\end{equation}
This simple derivation of the linear potential in the QCD-like model is 
impressive, because the Kaluza-Klein compactification that breaks supersymmetry plays an important role in the derivation. 
The mechanism for this appearance of linear potential is the same 
as that we have explained for the meson system in Sec.3; indeed, the 
string tension $k$ here is identical to $f(U_{\rm MIN})/2\pi$ in 
Eq.~(\ref{horizontal}). 
It is also notable 
that the dual gravity theory can give very straightforward by both the linear 
potential in this section and the Coulomb potential in the last section if the five-dimensional background curved spaces are properly chosen.

Next, we analyze the string segments in the second part that is relevant to the diquark system, $(J_1, (q_1q_2))$ and $(J_1', (q'_1q'_2))$.  Here, we have the following relations, similar to those given in the previous section: 
\begin{eqnarray}
E_2&=&\frac{U_1}{2\pi} \int^{U_2/U_1}_{1} dy \sqrt{ \frac{y^3}{y^3-A'^2}},  \label{QCDE2} \\
z_{(2), (3)}&=&\frac{(R'U_0)^{3/2}}{2(U_1)^2} \int^{U_2/U_1}_{1} dy \frac{1}{\sqrt{y^3(y^3-A'^2)}}, \label{QCDz2}
\end{eqnarray}
where $U_0 \ll U_1$ is assumed, so that 
\begin{equation}
A'^2=\frac{1}{4}\left(\frac{U_0}{U_1}\right)^3 +
\frac{3}{4} \approx \frac{3}{4}.
\end{equation}
In the limit of large extent for $U$, $U_1 \ll U_2$, we have
\begin{equation}
U_1^2=\frac{c'_2(R'U_0)^{3/2} }{2z_{(2),(3)}}+\cdots,
\end{equation}
where $c'_2$ is given by
\begin{equation}
c'_2=\int^{\infty}_{1} dy \frac{1}{\sqrt{y^3(y^3-3/4)}} \approx 0.628.
\end{equation}

Because from Eq. (\ref{xperp2}) we have 
\begin{equation}
|{\bf x}_{\perp(2), (3)}|=\sqrt{3}\left(\frac{U_1}{U_0}\right)^{3/2} z_{(2), (3)}, \label{QCDxperp}
\end{equation}
we obtain
\begin{equation}
U_1^{1/2}=\frac{\sqrt{3}}{2}c'_2(R')^{3/2} \frac{1}{|{\bf x}_{\perp(2), (3)}|}.
\end{equation}

Then, we have the expression 
\begin{equation}
E_2(U_2; |{\bf x}_{\perp(2), (3)}|)=\frac{1}{2\pi} \left( U_2- c'_3 \left( \frac{\sqrt{3}}{2} c'_2(R')^{3/2} \right)^2  \frac{1}{{|{\bf x}_{\perp(2), (3)}|}^2} \right) +\cdots, \label{QCDE_2'}
\end{equation}
where $c'_3$ is given by
\begin{equation}
c'_3=1-\int^{\infty}_{1} dy \left(\sqrt{\frac{y^3}{y^3-3/4}}-1\right)\approx 0.736.
\end{equation}
Like $c_2$ and $c_3$ considered about, $c'_2$ and $c'_3$ are not independent, related as
\begin{equation}
2c'_3-1=\frac{3}{4}c'_2.  \label{A3}
\end{equation}

In this QCD-like model, the relative potential inside the diquark system is not of the  Coulomb type in the large extent limit, but it is of the critical 
form of the collapsible potential $\propto1/(\mbox{distance})^2$. For this reason, it may be thought that this system would collapse. 
However, actually the string energy is positive definite, and thus 
the system is guaranteed to be stable from the beginning. 
To see this more explicitly, we now evaluate the energy in limit of a small extent for $U$, assuming $U_1 \le U_2$, but $U_1 \approx U_2$.

Expanding $y$ in a power series near $y=1$,  Eqs.(\ref{QCDE2}) and (\ref{QCDz2}) read
\begin{eqnarray}
E_2&=&\frac{1}{\pi}(U_2-U_1)+ \cdots,  \\
z_{(2), (3)}&=&\frac{(R'U_0)^{3/2}}{U_1^3}(U_2-U_1)+\cdots.
\end{eqnarray}
From Eqs.(94), (95) and (\ref{QCDxperp}), the energy for the three segments in the second part becomes
\begin{equation}
E_2=\frac{1}{\sqrt{3}\pi}\left(\frac{U_2}{R'}\right)^{3/2} |{\bf x}|_{(2), (3)}+ \cdots, 
\end{equation}
where we have used the condition $U_1 \approx U_2$.

Now, a linear potential appears near $|{\bf x}|_{(2), (3)} \approx 0$, 
and it can prevent the diquark system from collapsing.
The appearance of the linear potential in the small distance limit is trivial: Because the string energy $E(r)$ is a function of the distance $r$  
regular around $r=0$ and $E(r=0)=0$, the linear potential is nearly equal to 
the first term of the Taylor expansion.

From the above treatment, we obtain also in the QCD-like model, the following potential $V$, defined in Eq.(\ref{defPotential}):
\begin{equation}
V=k(z_1+z_2+z_3)+\sum_{i=1,2}\left(V_{\perp}(x_{i})+V_{\perp}(x'_{i})\right),
\end{equation}
where
\begin{equation}
V_{\perp}(x) \sim\left\{
\begin{array}{@{\,}ll}
\displaystyle -\frac{a}{M}\frac{1}{x^2} & \mbox{for}~~x \rightarrow\infty,  \\[2ex]
\displaystyle -\frac{1}{2}m +\frac{2}{\sqrt{3}} (m^3 M)^{\frac{1}{2}}x  & \mbox{for}~~x \rightarrow0. \end{array}
\right.  \label{Vperp}
\end{equation}
Here, using the identity (\ref{A3}), we have
\begin{eqnarray}
a&=&\frac{3}{4} (c'_3-1/2) (c'_2)^2=\frac{9}{32}(c'_2)^3\approx 0.0689, \\
M&=&M_{KK} /(\alpha_{c} N_{c}), 
\end{eqnarray}
where the  parameter $M$ is defined with the mass parameter $M_{KK}$ of the Kaluza-Klein compactification given in Eq.(\ref{MKK}).  The choice of coordinates $z_{i}$ $(i=1-3)$, and $(x_1, x_2)$ and $(x'_1, x'_2)$, is the same as in the last section.
In the above derivation of potential, $U_1$ appearing in the segments $(J_0, J_1)$ and $(J_0, J_1')$ has been rewritten in terms of physical variables.  In limit of large extent for $U$, we have used the averaged expression, as before, 
\begin{equation}
U_1=\left(\frac{\sqrt{3}}{2}c'_2\right)^2 R'^3 \frac{1}{2} \left( \frac{1}{x_{1}^{(\prime)2}}+\frac{1}{x_{2}^{(\prime)2}} \right),
\end{equation}
while in limit of small extent, we have used
\begin{equation}
\frac{1}{2\pi} U_1\approx \frac{1}{2\pi} U_2=  \frac{1}{2}(m_1+m_2).
\end{equation}
 
Next, we solve the Schr\"odinger equations and evaluate the lowest eigenvalues.  Here, the relevant potentials are the linear potential $V_{z}(z)=kz$ and $V_{\perp}(x)$ in Eq.(\ref{Vperp}).

To estimate roughly the lowest eigenvalues, we apply the limit of small extent for $V_{\perp}(x)$, since the collapsible potential at long distance is supported by the short distance contribution, and the lowest eigenvalue may depend strongly on the short distance behavior.  In this rough approximation, the problem is reduced to solving the Schr\"odinger equation with the linear potential.  To find the eigenvalue naively, we use the uncertainty principle.  This is because the lowest eigenvalue is  to be determined by balancing the attractive force from the classical potential and the repulsive force from the uncertainty principle.  Explicitly, we replace $(-i{\bnabla})^2$ by $1/r^2$ with our convention $\hbar=1$ and estimate the lowest eigenvalue $E_0$ as the stable value of the energy under the variation of $r$.  For a particle of mass $m$ moving in the potential $V(r)=k r^{n}$, we obtain this $E_0$ as
\begin{equation}
E_0 \approx \frac{1}{2}(n+2) k^{\frac{2}{n+2}} (nm)^{-\frac{n}{n+2}},
\end{equation}
which gives 
\begin{equation}
E_0 \approx \frac{3}{2}k^{\frac{2}{3}}m^{-\frac{1}{3}}
\end{equation}
in the linear potential case, $n=1$.

Even with such naive considerations, we obtain, similarly to the previous section, the following mass formula for pentaquarks:
\begin{eqnarray}
&&\bar{M}(\mbox{pentaquark}) \nonumber \\
&=&\bar{m}_1+\bar{m}_2+\bar{m}_3  \nonumber\\
&+&\frac{2}{3}(\alpha_{c} N_{c})^2 \{ 2 (\bar{m}_1+\bar{m}_2)^{-\frac{1}{3}} +(\bar{m}_3)^{-\frac{1}{3}} \} \nonumber \\
&+&  6^{\frac{2}{3}} \{(\bar{m}_1)^{\frac{2}{3}}+(\bar{m}_2)^{\frac{2}{3}} \}.
\end{eqnarray}

All the masses with bars are dimensionless constants normalized by $M$. Note that the first two terms in (105) are $\bar{m}_1+\bar{m}_2$, instead of $2(\bar{m}_1+\bar{m}_2)$. This is because the negative contribution $-(\bar{m}_1+\bar{m}_2)$ comes from the constant term $-(1/2)m$ in $V_\perp$ in Eq.(98):
 
\begin{eqnarray}
& &\bar{M}(\mbox{triquark}) \nonumber \\
&=&\bar{m}_1+\bar{m}_2+\bar{m}_3  \nonumber\\
&+&\frac{2}{3}(\alpha_{c} N_{c})^2 \sum_{i=1-3} (\bar{m}_{i})^{-\frac{1}{3}} .
\end{eqnarray}
Here, the pentaquark and triquark are considered to be $((q_1 q_2)^2 \bar{q}_3)$ $(q_1 q_2 q_3)$, respectively. 

As we know, the mass formulae derived above is too naive, adopting various rough approximations, and therefore it seems inappropriate to  compare these formulae with experimental data.  Nevertheless, we carry out such a comparison here. 

The input parameters are $M(N)=$939 MeV and $M(\Sigma)=$1,193 MeV, using averaged masses among their charge multiplets.  Then, fixing  $\alpha_c$ to 0.33 (or $N_c \alpha_c=1$), the quark masses and the pentaquark masses are estimated respectively as $m_u=m_d=$313--312 MeV, $m_s=$567--566 MeV, and  $M(\Theta)=$1,577--1,715 MeV, $M(\Xi)=$1,670--1,841 MeV, corresponding to the KK mass scale of $M=M_{KK}=$2--5 MeV.

If we choose the  heavier value $M=M_{KK}=$10 MeV, then the pentaquark masses become  heavier  than expected, $M(\Theta)=$1,852 MeV and  $M(\Xi)=$2,010 MeV.   Therefore, to approach the observed values of the pentaquarks, we have to use a slightly larger extra dimension with radius $R_{KK}=2\pi/M_{KK}=$2--5 MeV,  several hundred femto meters.  

Here  again, we can say that the results obtained do not differ significantly from experimented values, ever in the extremely naive approximations adopted in our study. Therefore, we hope to obtain better predictions if we can make the treatment of pentaquarks in the string picture more rigrous. There is room for such improvement.

\section{Production and decay of pentaquarks}
In this section we study the production and decay of pentaquarks by examining how the decay rate can be estimated.

The key issue cocerning pentaquarks is whether or not we can predict the extremely narrow width, which is probably around 1 MeV.
Production is the inverse process of decay, and therefore it is sufficient to examine the decay process $\Theta^{+} \to n+K^{+}$ in the string picture.  We display the possible decay processes in Fig.6.  The main step is the recombination of two strings.  There are two channels; in the first channel, the string segments $(J_0, \bar{s})$ and $(u, J_1)$ or $(u', J'_1)$ are recombined, while in the second channel, $(J_0, J_1)$ and $(d', J'_1)$, [or  $(J_0, J'_1)$ and $(d, J_1)$] are recombined. The recombination of strings is the process given described in Fig.7.

\begin{figure}
\begin{center}
\includegraphics[width=15cm,clip]{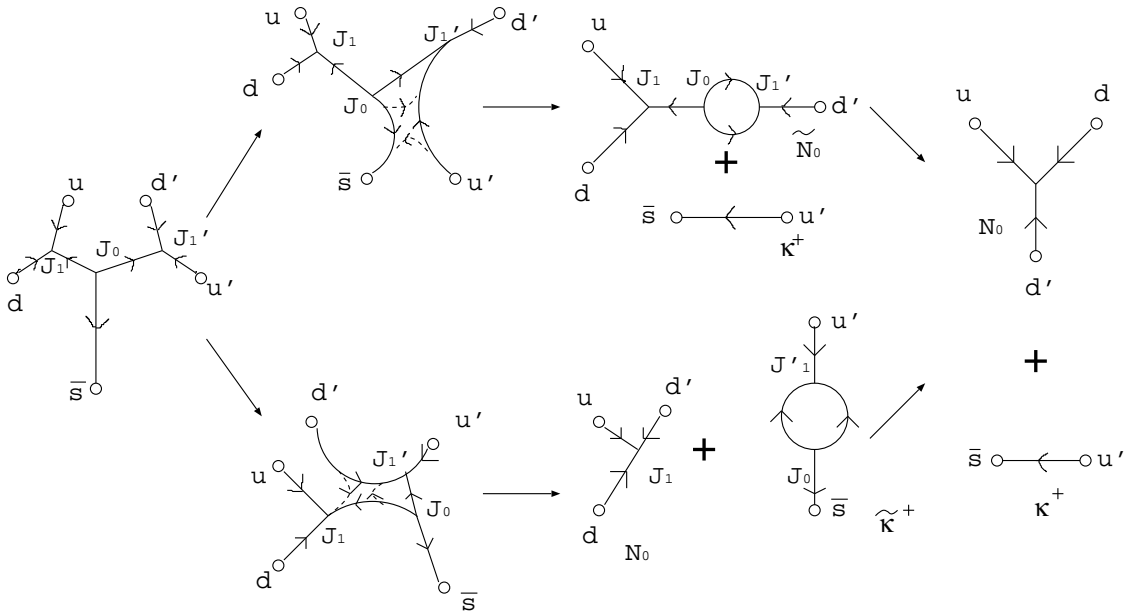}
\end{center}
\caption{Decay processes of $\Theta^+ \to K^{+} + N^0$ (neutron).}
\end{figure}%

\begin{figure}
\begin{center}
\includegraphics[width=5cm,clip]{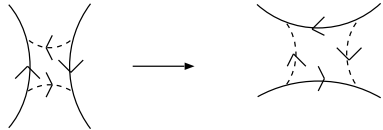}
\end{center}
\caption{Recombination of strings.}
\end{figure}%

 In the first channel, neutron  $n$  (denoted hereafter as $N^0$, the neutral component nucleon) accompanied by a ``string loop" is created, while in the second channel, a $K$ meson with the string loop is created.  We denote these excited states as $\tilde{N}^0$ and $\tilde{K}^{+}$.  They are not uniquely determined, as they depend on the shape and size of the string loop, but we denote them in general as $\tilde{N}^0$ and $\tilde{K}^{+}$. Then, two decay channels can be written as follows:
\begin{equation}
\Theta^{+} \rightarrow\left\{
\begin{array}{@{\,}ll}
& \tilde{N}^{0}+K^{+}\to N^0+K^{+}~~ \mbox{(channel 1)}, \\
& N^0+\tilde{K}^{+} \to N^0+K^{+}~~ \mbox{(channel 2)}.
\end{array}
\right.
\end{equation}

In addition to recombination, different processes are possible to produce $\tilde{N}^{0}$ and  $\tilde{K}^{+}$.   The process that produces $\tilde{N}^{0}$ is pair production of $(u''\bar{u}'')$ through string splitting on the segment $(J_0, \bar{s})$ and the subsequent pair annihilation of $(\bar{u}'' u)$ or $(\bar{u}'' u')$. That producing $\tilde{K}^{+}$ is the pair production of $(d''
\bar{d}'')$ through string splitting on the the segment $(J_0, J_1)$ [or $(J_0,
J'_1)$], and the subsequent pair annihilation of $(\bar{d}'' d')$ [or
$(\bar{d}'' d)$].   We include these different processes into the preliminary stages of the above mentioned two channels. 

Now, we present the following effective Lagrangian (1) to control the decay process of pentaquarks:
\begin{eqnarray}
{\cal L}_{(1)} &=&\tilde{m}\bar{\tilde{N}} N+\frac{g_A}{f_{K}}~ \bar{\Theta}\gamma^{\mu}\gamma_5 \tilde{N}\partial_{\mu}K \nonumber \\
&+& \tilde{m}'^2 \tilde{K}^{\dagger} K+\frac{g'_A}{f'_{K}} ~\bar{\Theta}\gamma^{\mu}\gamma_5 N\partial_{\mu}\tilde{K}.
\end{eqnarray}
Here, $g_A$ and $g'_A$ are the relevant axial couplings, and $f_K$ and $f'_{K}$ are the decay constants of $K^{+}$ and ${\tilde K}^{+}$.

Here, we have used the PCAC hypothesis or the dynamical generation of the $K$ meson as a Nambu-Goldstone mode of the bound state ~\cite{Nambu}.  Of course, we have to re-examine this dynamical problem further in the string picture, in which the bound states are easier to analyze than in the usual field theory, while the currents are more difficult to describe, and the ``currents" are non-local in the present case, including  those of the gauge theory and those of the gravity theory.  Referring to our review of the dual gravity theory of QCD given in Sec.2, we are tempted to determine the correspondence of ``currents" between gauge theory and its dual gravity theory.  

In any case, we can examine the decay $\Theta^{+} \to N^0+K^{+}$ using the effective Lagrangian (1).
Taking account of the two channels, we estimate the $\Theta NK$ coupling $g_{\Theta NK}$ in the effective Lagrangian (2) as
\begin{equation}
{\cal L}_{(2)}=g_{\Theta NK} \bar{\Theta}\gamma_5 NK.
\end{equation}
As a result, the coupling  $g_{\Theta NK}$ is determined by 
\begin{eqnarray}
g_{\Theta NK}&=&2g_A \left(\frac{M(\Theta)+M(\tilde{N})}{M(\tilde{N})-M(N)} \right)\left(\frac{\tilde{m}}{f_{K}}\right) \nonumber \\
&+& 2g'_A \left( \frac{M(\Theta)+M(N)}{M(\tilde{K})^2-M(K)^2} \right) \left( \frac{(\tilde{m}')^2}{f'_{K}} \right).
\end{eqnarray}

Usually, the $\pi$-nucleon coupling $g_{NN\pi}$ is $O(10)$, giving the standard decay width of $O(100)$ MeV for the nucleon family. Therefore, in order to obtain a narrow width of $O(1)$ MeV for the pentaquarks, we have to require $g_{\Theta NK} \sim O(1)$.  Then, we estimate roughly the mass mixing parameters of $N^0$ and  $\tilde{N}^{0}$ and of $K^{+}$ and $\tilde{K}^{+}$ as follows:
\begin{eqnarray}
\tilde{m} &\sim&f_{K} \left( \frac{M(\tilde{N})-M(N)}{M(\Theta)+M(\tilde{N})} \right) \sim100~~\mbox{MeV}, \\
\tilde{m}' &\sim& \sqrt{ 
f'_{K} \left( \frac{M(\tilde{K})^2-M(K)^2}{M(\Theta)+M(N)} \right)} \sim100~~\mbox{MeV}.
\end{eqnarray}
These mass mixings are roughly $1/10$ as large as their masses.  This implies that the formation of a ``string loop" is rather strongly suppressed.

The properties studied above using the effective Lagrangians should be examined within the string picture.  We leave this problem to a future study.


\section{Preliminary sketch of spin, hyperfine interaction and parity}
In this section, we give a preliminary sketch of how spin, the hyperfine interaction and parity of state (such as a pentaquark) can be understood in the string picture.  For this purpose we first recall the quark model and its prediction of hadron masses.

In the quark model, the prediction of hadron masses is in good agreement with their experimental values if the masses of the hadrons are identified with the sum of the constituent quark masses and the hyperfine interactions between them~\cite{HF}.  That is, the hadron mass $M$ is expressed as follows:
\begin{eqnarray}
M&=&\sum_{i} m_{qi} \nonumber \\
&-& \frac{8\pi}{3} \alpha_{s} \sum_{i>j}\frac{1}{m_i m_j} \langle \frac{{\bf \lambda}^{c}_i}{2} \cdot \frac{{\bf \lambda}^{c}_j}{2}\rangle \langle \frac{{\bf \sigma}_i}{2} \cdot \frac{{\bf \sigma}_j}{2}\rangle |\psi(0)|^2, \label{HF1}
\end{eqnarray}
where ${\bf \lambda}^{c}_i$ is the Gell-Mann matrix of color $SU(3)_{c}$ for the $i$-th quark.  The second term represents the ``hyperfine interactions" due to the color magnetic forces, being calculated from a one-gluon exchange process.  Even if we use the ordinary values of the constituent quark masses $m_{u, d}=$ 360 MeV and $m_{s}=$ 540 MeV and include the hyperfine interactions of color magnetic forces, the predictions of  $M(\Theta^{+})$ and $M(\Xi^{--})$
seem to be overestimated by 100--200 MeV.  That is, some extra attractive force is missing in the quark model.~\cite{Oka}   In this respect, the study presented in this paper is useful, as it can give various extra forces coming from non-perturbative effects in the string picture and going beyond the one-gluon exchange process.  For example, linear potential appears directly from the QCD like model.  Therefore, our study may fill the gap existing between the quark model and the experimental data.

To this point, we have not mentioned the effects of ``spin".  As is well known, the spin matrix is defined in terms of $\gamma$ matrices as 
\begin{equation}
\frac{\sigma^{\mu\nu}}{2}= \frac{i}{4}[\gamma^{\mu}, \gamma^{\nu}].  \label{Defspin}
\end{equation}
The spatial components of $\sigma^{\mu\nu}$ give the spin matrices, $\frac{\sigma^{12}}{2}=\frac{{\bf \sigma}^3}{2}={\bf s}^3$, {\it etc}.

In superstrings, we have fermionic variables on the world sheet, $\psi^{\mu}(\tau, \sigma)$, in addition to the bosonic variables $X^{\mu}(\tau, \sigma)$ specifying the location of the world sheet. The equal-time commutation relation of the former reads 
\begin{equation}
\{ \psi^{\mu}(\tau, \sigma), \psi^{\nu}(\tau, \sigma') \}=\eta^{\mu\nu}\delta(\sigma-\sigma'),
\end{equation}
whose zero mode relation, $\{ \sqrt{2}\psi_0^{\mu}, \sqrt{2}\psi_0^{\nu}\}=2\eta^{\mu\nu}$, is identical to the commutation relation of gamma matrices.  Therefore, the fermionic variable, $\psi^{\mu}(\tau, \sigma)$, in string theory is a distribution function of $\gamma$ matrices on the world sheet.

In superstring theory, ``spin" is defined as a generalization of (\ref{Defspin}) by
\begin{equation}
\frac{\sigma^{\mu\nu}}{2}(\tau, \sigma)= \frac{i}{2}[\psi^{\mu}(\tau, \sigma), \psi^{\nu}(\tau, \sigma)]=i\psi^{\mu}(\tau, \sigma) \psi^{\nu}(\tau, \sigma).
\end{equation}
This shows that ``spin" is not a constant matrix in string theory but, rather, a density distribution function on the world sheet, parametrized by $(\tau, \sigma)$.  \footnote{The spin distribution function also gives oscillation modes, or spin waves on the world sheet.}  

Because the short distance behavior of a string is soft, as is well known from the disappearance of ultraviolet divergences in string theory, the following problem arises: Is the hyperfine contact interaction in Eq.(\ref{HF1}) consistent with string theory?

To answer this question, we study in the following the hyperfine interaction in string theory.

In string theories, quantum numbers such as ``color" and ``flavor" are incorporated at the endpoints of an open string.  However, spin is not located at the endpoints, but is distributed on the whole world sheet of $(\tau, \sigma)$ plane.  
This is consistent with the fact that the gravitational interaction that occurs in the middle of strings can feel spins.  Only quantum numbers irrelevant to the gravitational interaction can be placed on the boundary of strings. 
Here, a new concept of quantum numbers in the string picture emerges: 
Colors and flavors are located at the endpoints of the string, but spins are distributed over the whole string.

First, we give naive discussion of how the masses of spin $\frac{1}{2}$ states such as pentaquarks can be predicted.  In the supersymmetric case of Maldacena's prototype model, the spin $0$ state and the spin $\frac{1}{2}$ state are degenerate, and therefore the mass formulae for spin $\frac{1}{2}$ pentaquarks are those obtained in this paper without considering any spins.

However, if supersymetry is broken, as in the QCD-like model, we cannot rely on such a simple treatment.  Nevertheless, if the effect of supersymmetry breaking is reduced to only the anti-periodicity of fermions in the compactified direction $\vartheta$, it is sufficient to add the extra KK energy $\frac{1}{2} M_{KK}$ to the mass formulae of the QCD-like model in Sec.5. In this case,  we have a spin 0 family of pentaquarks that is a few MeV lighter, because $\frac{1}{2} M_{KK}$ is as small as a few MeV.  These simple arguments may reach a deadlock.

In order to take account of the ``spin" more seriously, let us go back to the work of Nambu in 1970~\cite{Nambu2}.  At that time, he considered the question of how to incorporate the quantum numbers of $SU(3)'\times SU(3)''\times SU(2)$ (Han-Nambu model of three triplets, or being allowed to replace them by flavor and color in the present context) $\times$ (Dirac spin) in  string models.  In this first paper of applying strings to particle physics,  we can fortunately find a clue to solving our problem.  

First, consider Eqs. (39) and (40) of that paper.  They read
\begin{eqnarray}
I&=& g \sum_{(n, n')} \left(\gamma^{(n)}_{\mu} \gamma^{\mu(n')} + \mbox{const}\right),  \label{N1} \\
I'&=& \sum_{n} \gamma^{(n)}_{\mu} \phi^{\mu(n)}. \label{N2}
\end{eqnarray}
Here, $\gamma^{\mu}$ is the ordinary $\gamma$ matrix (which can be regarded
as $\gamma_5\gamma^{\mu}$). Such matrices are distributed on the world sheets, as explained above, and they experience the interaction between the nearest neighbor sites $\{(n, n')\}$ in (\ref{N1}).  In the second action, (\ref{N2}), the ``spin" couples to an external magnetic field $\phi$.  Nambu regarded these as the magnetic interaction of a two-dimensional anti-ferromagnet and an ion crystal.

Eqyation (\ref{N1}) is identical to the free fermionic action of a string, and so Nambu's string theory is from the beginning the spinning string of Neveu-Schwarz and Ramond.   Equation (\ref{N2}) is extremely suggestive for our problem.

It is now not difficult to study the hyperfine interaction in string theory.  
Equation (\ref{N2}) can be imported into our string model as follows~\cite{Endo}: 
\begin{eqnarray}
S_2(\mbox{NS-NS+gluon})&=&\int d\tau d\sigma \psi^{\mu}(\tau,\sigma)
\psi^{\nu}(\tau,\sigma) M_{\mu\nu}(X(\tau, \sigma)), \\
 \mbox{with}~~M_{\mu\nu} &=& G_{\mu\nu}+N_{\mu\nu}+B_{\mu\nu}+g_{c}
\frac{\lambda^c}{2}F^c_{\mu\nu},
\end{eqnarray}
where the graviton, $G_{\mu\nu}$, the Kalb-Ramond field, $B_{\mu\nu}$, the field strength of the branes' deformation modes, $N_{\mu\nu}$, and the field strength of gluons, $F^c_{\mu\nu}$, couple to ``spins" distributed on the world sheet ``magnetically".
In addition, we have 
\begin{equation}
S_2(\mbox{R-R})=\int d\tau d\sigma \psi^{\mu}(\tau, \sigma)
\psi^{\nu}(\tau, \sigma) \prod_{\lambda\ne \mu, \nu} \psi^{\lambda}(\tau,
\sigma) \times A^{(2)}_{\mu\nu}(X(\tau, \sigma)) +\cdots,
\end{equation}
where various $n$-form R-R fields $A^{(n)}$ couple to the spins on the world sheet.  Contracting the ``magnetic fields" and
generating the propagation of the closed strings, we obtain the hyperfine interactions between spins on the world sheet.

In carring out the contraction, the contribution of NS-NS fields (attractive force) and that of the R-R fields (repulsive force) may cancel, in a similar manner as in the cancellation of the forces between two D-branes, as reviewed in Sec.2.  Therefore, it may be the case that only the contribution from gluons remains. 
As a result, we obtain the following:
  \begin{eqnarray}
& &S(\mbox{stringHF}) \hfill \nonumber \\
&\propto&\frac{-i}{2} \int d\tau d\sigma\int d\tau' d\sigma'
\psi^{\mu}(\tau, \sigma) \psi^{\nu}(\tau, \sigma) \psi^{\lambda}(\tau',
\sigma') \psi^{\rho}(\tau', \sigma') \nonumber \\
&\times& g_{c}^2\langle \frac{\lambda^a}{2} F^a_{\mu\nu}(X(\tau, \sigma))
\frac{\lambda^{b}}{2}F^{b}_{\lambda\rho}(X(\tau', \sigma'))\rangle,
\end{eqnarray}
where the propagator of the field strength gives the contact interaction proportional to $\delta^{(D)}(x-x')$, where {\it D} is the number of dimensions in which gluons can propagate.  

Then, we obtain the hyperfine interaction in string theory for the fixed 
color and spin states $|c, s\rangle=|c\rangle|s\rangle$ as follows:
\begin{eqnarray}
& &
\langle c, s|S(\mbox{stringHF})|c, s\rangle 
\propto 8\pi\alpha_c \langle c| 
\frac{\lambda^a}{2}\frac{\lambda^a}{2}|c\rangle \hfill
\nonumber \\
&\times& \int {\cal D} X^{\mu} (\tau, \sigma) \int {\cal D} \psi^{\mu} (\tau, \sigma) \int d\tau d\sigma \int d\tau' d\sigma' 
\langle s|\frac{\sigma_{\mu\nu}}{2}(\tau,\sigma)
\frac{\sigma^{\mu\nu}}{2}(\tau', \sigma')|s\rangle
\nonumber \\
&\times& \delta^{(D)}(X^{\mu}(\tau, \sigma)-X^{\mu}(\tau, \sigma'))|
\Psi_{s, c}[X^{\mu}(\tau, \sigma), \psi^{\mu}(\tau, \sigma)] |^2.
\end{eqnarray}
Here, $\Psi_{s, c} \left[X^{\mu}(\tau, \sigma), \psi^{\mu}(\tau, 
\sigma)\right]$ is the component of
the wave functional of the string state (such as pentaquark) at a fixed
representation of color and spin.  In other words, the wave functional of the string is expanded in terms of  color $|c\rangle$ and spin $|s\rangle$ states as 
follows:
\begin{equation}
\Psi[X^{\mu}(\tau, \sigma), \psi^{\mu}(\tau, \sigma)]=\sum_{s, c} \Psi_{s, 
c} \left[X^{\mu}(\tau,
\sigma), \psi^{\mu}(\tau, \sigma)\right] |s\rangle |c\rangle.
\end{equation}
In this expression, $\delta^{(D)}(X^{\mu}(\tau,
\sigma)-X^{\mu}(\tau',
\sigma'))\left| \Psi_{s, c} \left[X^{\mu}(\tau, \sigma), \psi^{\mu}(\tau, \sigma)\right] \right|^2$
appears. This gives the probability of the appearance of the configuration
$X^{\mu}=X^{\mu}(\tau, \sigma)$, in which two points on the world sheet
are
stuck together.
This is the string version of the factor $|\psi({\bf x=0})|^2$ in
Eq.(\ref{HF1}), where the locations of the two quarks are identical.

It is a challenging problem to evaluate this hyperfine interaction (which may or may not be negligible) in our string picture and to include the effect in the pentaquark mass formulae, where we need to know the wave functional of pentaquarks.   

Another issue is the parity of pentaquarks, which to this time has not been confirmed by experiments.  Theoretically, to determine the party of the lowest energy state is strongly related to the question of where the spins are located.   

In the usual picture of the quark model, spins are located at the endpoints, the same places at which colors and flavors are located, so that the symmetry properties determine the relative angular momentum between diquarks.  In the diquark model of~\cite{JW}, as an example, the angular momentum of two diquarks becomes a P-wave, and we obtain even parity for the pentaquarks.  
However, in our string picture, spins are not localized, but seem to be distributed over the seven string segments of the pentaquarks, similarly to the manner in which as spins are distributed on an anti-ferromagnet or ion crystal webs.  Therefore, the treatment of parity in the string picture may differ from that in 	the ordinary quark model.  This is also an open problem for future study.

\section{Conclusion}
In this paper we have studied the recently observed pentaquark baryons $\Theta^{+}$ and $\Xi^{--}$ in the dual gravity theory of QCD. First, we reviewed this dual gravity theory of QCD, using the familiar concepts of factorization and vacuum insertion.  Next, we developed a general formulation which is useful to study the branched string web (such as the string picture of the pentaquark) located in a curved background space.  In this formulation, we derived  simple connection conditions to be imposed at the branched points (or junctions) of the web.

Using this general formulation, mass formulae of the pentaquark baryons were easily derived in Maldacena's prototype model ($AdS_5 \times S^5$) for a supersymmetric QCD. Mass formulae were also derived in a more realistic model  ($AdS$-Scwartzshild geometry $\times S^4$) for ordinary QCD.  In the former model, we prepared the stuck $N_c$ colored D3-branes  and $N_f$ flavored D7-branes located separately, according to the intrinsic masses of quarks with flavors.  In the latter, model  the $N_c$ colored D4-branes  and $N_f$ flavored D6-branes were prepared.

Even with such naive approximations, the predictions of the pentaquark masses in both models do not differ significantly from their experiment values.  For example, in the QCD-like model, we obtain the quark masses and pentaquark masses as follows: 
$m_u=m_d=$313--312 MeV, $m_s=$567--566 MeV, and  $M(\Theta)=$1,577--1,715 MeV, $M(\Xi)=$1,670--1,841 MeV, for a KK mass scale of $M=M_{KK}=$2--5 MeV.

The decay processes of pentaquarks were also examined considering the recombination of strings, for which we give a discussion on how we can obtain the decay width of $O(1)$ MeV. 

With the above results, we can conclude that the present approach is promising for the purpose of predicting the observed values of the pentaquark masses and accounting for the narrow decay widths.

Throughout this paper, we ignored the effects of spin.  For a more reliable study of pentaquarks, we gave a sketch of how spin, the hyperfine interaction and parity are taken into account in the string picture of hadrons.  
The sketch is still preliminary, but the new concept regarding the location of color, flavor and spin quantum numbers, according to which, ``color and flavor are located at the endpoints of the string, and the spin is distributed over the whole string, arises. This concept may give new insight into hadron physics.

%
%
\section*{Acknowledgements}
The authors are grateful to Shigeki Sugimoto for useful discussions on string theories, especially on the AdS/CFT correspondence, to Teiji Kunihiro for various information on pentaquarks, and to Yukiko Ohtake for a valuable comment. We give our thanks to the proofreader for his improving the English.

M. B., T. K and A. S. are partially supported by Grants-in-Aid for Scientific Research (Nos.12047225, 16340071 and 14039204, respectively), from the Ministry of Education, Culture, Sports, Science and Technology, Japan.

\appendix
\section*{Appendix}

Here, we give a proof of some of the identities used in this paper.
First, we give a proof of the identity $c_1+\tilde c_1 =1$ 
appearing in Eq.(\ref{A1}).  More explicitly, this identity reads
\begin{equation}
\int^{\infty}_{1} dy \left(
{y^2 \over \sqrt{y^4-1}}-1 \right) + \int^{\infty}_{1} dy 
\frac{1}{y^2\sqrt{y^4-1}}=1.
\end{equation}

We define
\begin{equation}
C \equiv {y^2 \over \sqrt{y^4-1}}, \quad
D \equiv {1 \over y^2\sqrt{y^4-1}}.
\end{equation}
Then, we clearly have
\begin{equation}
C+D = {d\over dy}\left({\sqrt{y^4-1}\over y}\right).
\end{equation}
Subtracting 1 from both sides and integrating over $y \in [1, Y]$
we have
\begin{equation}
\int_1^Y dy\,(C-1) +
\int_1^Y dy\,D = \left[ {\sqrt{y^4-1}\over y}-y\right]_1^Y
= 1 + Y\left(\sqrt{1-Y^{-4}}-1\right),
\end{equation}
which gives the desired identity, $c_1+\tilde c_1 =1$, if we take $Y\to \infty$.

In the same way, we can prove the following identities:
\begin{eqnarray}
\int^{\infty}_{1} dy \left(
{y^2 \over \sqrt{y^4-A^2}}-1 \right)
+A^2 \int^{\infty}_{1} dy \frac{1}{y^2\sqrt{y^4-A^2}}
&=&1-\sqrt{1-A^2}, \\
\int^{\infty}_{1} dy \left(
\sqrt{{y^3 \over y^3-A'^2}}-1 \right)
+\frac{A'^2}{2} \int^{\infty}_{1} dy \frac{1}{\sqrt{y^3(y^3-A'^2)}}
&=&1-\sqrt{1-A'^2}.
\end{eqnarray}
If we take $A^2={3 \over 4}$ in the first equation, we have $2c_3-1={3 \over 2} c_2$ in Eq.(\ref{A2}), while if we take $A'^2={3 \over 4}$ in the second 
equation, we have $2c'_3-1={3 \over 4} c'_2$ in Eq.(\ref{A3}).

The integrals in the above expressions are elliptic integrals, and their 
inverse functions in terms of $Y$ are elliptic functions.  Therefore, the 
constants $c_1, c_2 and c_3$ are the values at which the elliptic functions are infinite.

It is interesting that these constants appear 
as the coefficients of the effective potential in the large separation 
limit. Therefore they are related to the coefficient functions of the renormalization group equations 
on the low energy side.  This fact is to be clarified from the relationship between string theory and elliptic 
functions and that between the renormalization group equation and five-dimensional field equations, with the energy scale $u$ as the fifth coordinate.

%
%

\end{document}